\documentclass[a4paper,11pt]{article}

\usepackage{jheppub} 
\usepackage{physics}

\usepackage[T1]{fontenc} 

\preprint{OU-HET 1052}

\title{\boldmath 
Out-of-time-order correlator \\
in coupled harmonic oscillators
}

\author{Tetsuya Akutagawa,}
\author{Koji Hashimoto,}
\author{Toshiaki Sasaki,}
\author{Ryota Watanabe}

\affiliation{Department of Physics, Osaka University,\\1-1 Machikaneyama, Toyonaka, Osaka 560-0043, Japan}

\emailAdd{akutagawa@het.phys.sci.osaka-u.ac.jp}
\emailAdd{koji@phys.sci.osaka-u.ac.jp}
\emailAdd{sasaki@het.phys.sci.osaka-u.ac.jp}
\emailAdd{watanabe@het.phys.sci.osaka-u.ac.jp}

\abstract{
Exponential growth of thermal out-of-time-order correlator (OTOC) is an indicator of a possible gravity dual, 
and a simple toy quantum model showing the growth is being looked for.
We consider a system of two harmonic oscillators coupled nonlinearly with each other, and  
numerically observe that the thermal OTOC grows exponentially in time.
The system is well-known to be classically chaotic, and is a reduction of Yang-Mills-Higgs theory.  
The exponential growth is certified because the growth exponent (quantum Lyapunov exponent) of the thermal OTOC
is well matched with the classical Lyapunov exponent, including their energy/temperature dependence.
Even in the presence of the exponential growth in the OTOC, the energy level spacings 
are not sufficient to judge 
a Wigner distribution, hence the OTOC is a better indicator of quantum chaos.
}

\begin{document}

\maketitle
\flushbottom

\section{Introduction}
\label{sec:intro}

The out-of-time-order correlator (OTOC) \cite{Larkin} is considered as an indicator of quantum chaos.
As it is a quantum analogue of the classical sensitivity against tiny changes in initial conditions, 
the exponential growth of the OTOC in its time evolution is regarded as the indicator of the chaos.
The popular traditional indicator is the statistics of the energy level spacings 
(see for example \cite{QSC}), and the interest is
in whether the OTOC can be a better indicator of quantum chaos or not.

The OTOC attracted attention also because it can be an indicator of possible gravity dual, through the renowned AdS/CFT correspondence \cite{Maldacena:1997re}. Gedanken experiments about shock waves in black hole geometries 
\cite{Shenker:2013pqa,Shenker:2013yza} have led to a 
maximum bound \cite{Maldacena:2015waa} of the quantum Lyapunov exponent, the exponent showing up in thermal OTOCs. 
The saturated value is interpreted as the red shift near the event horizon of a black hole with the Hawking temperature.
The quantum Lyapunov exponent of the thermal OTOC can discriminate whether the system allows a gravitational picture.
Theories with any gravity dual are strongly quantum, so we need a quantum analogue of the Lyapunov exponent
as a discriminator, which is now provided by the thermal OTOC.

It was discovered \cite{Kitaev-talk-KITP} that 
the Sachdev-Ye-Kitaev (SYK) model \cite{Sachdev:1992fk,Kitaev-talk}, a quantum mechanical model of Majorana fermions,
develops the exponential growth in its thermal OTOC and the quantum Lyapunov exponent saturates the bound.
Stimulated by this discovery, research on OTOCs in various quantum systems have been carried out.
Generic scheme for measuring the thermal OTOC in quantum mechanics was provided in \cite{Hashimoto:2017oit}
and it was shown that the thermal OTOC for a quantum stadium billiard, a typical chaotic system, does not grow exponentially.
With adequate quantum states prepared, the expectation value of the OTOC was observed to grow exponentially 
for a kicked rotor \cite{Rozenbaum:2017zfo}, a stadium billiard \cite{Rozenbaum:2019kdl}, and the Dicke model \cite{Chavez-Carlos:2018ijc}, 
among few-body systems\footnote{For systems with large number of
degrees of freedom, see for example \cite{Shen:2016htm,Bohrdt:2016vhv,Bianchi:2017kgb,Lin:2018tce,Rammensee:2018pyk,Lin:2018luj,Wang:2019vjl,Hartmann:2019cxq,Dag:2019yqu,Borgonovi:2019mrk,Yan:2019wio}.}. 
However, these examples are not with the thermal OTOCs. 
The thermal OTOC is indispensable for the identification with a gravity dual,
hence a simple quantum system whose thermal OTOC grows exponentially is demanded.

In this paper, we study one of the simplest models: two harmonic oscillators coupled nonlinearly with each other,
\begin{align}\label{CHO}
	H = \left( p_x^2 + \frac{\omega^2}{4}x^2 \right) + \left( p_y^2 + \frac{\omega^2}{4} y^2 \right) + g_0\,  x^2 y^2  \, ,
\end{align}
and show that the thermal OTOC of this model grows exponentially.
Harmonic oscillators form a fundamental basis for any quantum field theory, which shows the importance of this
simple system. 
In numerous examples, nonlinear couplings between harmonic oscillators provide classical chaos, needless to mention
the popular example of the double pendulum. 
In our coupled harmonic oscillator (CHO) model,
the chosen coupling $x^2 y^2$ with a coupling constant $g_0$ is significant by the following three reasons.
First,
the coupling $g_0 x^2 y^2$ is a part of the BFSS matrix theory \cite{Banks:1996vh} which is dual to a quantum gravity in 
the large $N$ limit. As the coupling is a dimensional reduction of the Yang-Mills coupling, the model \eqref{CHO} serves as
the starting point of a road to the AdS/CFT correspondence\footnote{Models similar to \eqref{CHO} appear
in the context of the AdS/CFT correspondence and string theory matrix models, see \cite{Asano:2015eha, Hashimoto:2016wme, Berenstein:2016zgj, Akutagawa:2018yoe}. Classical chaos 
of the BFSS Matrix Theory has been analyzed in \cite{Gur-Ari:2015rcq,Berkowitz:2016znt} and quantum corrections are discussed in \cite{Buividovich:2018scl}.
Relatedly, universal chaotic behavior near black hole horizons was found \cite{Hashimoto:2016dfz} and studied in \cite{Morita:2018sen, Dalui:2018qqv, Hashimoto:2018fkb, Zhao:2018wkl, Morita:2019bfr, Dalui:2019umw}.
}. 
Second, this model \eqref{CHO} has attracted attention for long years 
\cite{Matinyan Savvidy Ter-Arutunian Savvidy (1981a), Matinyan Savvidy Ter-Arutunian Savvidy (1981b), Savvidy (1984),Chaos and Gauge Theories} due to its classical chaos. 
It is one of the most popular chaos models, with known classical Lyapunov exponent and quantum level statistics 
\cite{Pullen Edmonds (1981), Haller Koppel Cederbaum (1984)}.
Third, the model is obtained by a reduction from $SU(2)$ Yang-Mills-Higgs theory 
which is a basis of the Standard Model
of elementary particles\footnote{The thermal OTOC and its Lyapunov exponent for Hermitian matrix $\phi^4$ quantum field theory in 4 spacetime dimensions, at weak coupling, were calculated in \cite{Stanford:2015owe}. For semiclassical analyses of OTOCs in quantum mechanics, see \cite{Jalabert:2018kvl}.}, so the model \eqref{CHO} is fundamentally related to the actual phenomena in the universe\footnote{
The extension to lattice Yang-Mills theories \cite{Muller:1992iw,Biro:1993qc} was applied to rapid thermalization 
of heavy ion collisions \cite{Kunihiro:2010tg}.}.

We carefully demonstrate the numerical calculations of the thermal OTOC of the coupled harmonic oscillators \eqref{CHO},
and see the exponential growth, by identifying the growth rate (named ``quantum Lyapunov exponent'') with the
classical Lyapunov exponent of the system. This quantum-classical correspondence in the thermal OTOC, shown in the
simple Hamiltonian system \eqref{CHO}, may be extended to various other models with classical chaos, which will 
provide a novel arena for measuring quantum chaos.

This paper is organized as follows. In Sec.~\ref{sec:review1}, we review the classical chaos of the coupled harmonic oscillator model \eqref{CHO}. Sec.~\ref{sec:review2} is a review of \cite{Hashimoto:2017oit} summarizing
the numerical methods for calculating the thermal OTOC. Then in 
Sec.~\ref{sec:exponential growth of OTOC}, we numerically evaluate the thermal OTOC of the system \eqref{CHO}
and show that it grows exponentially in time. The quantum Lyapunov exponent measured there has the temperature
dependence expected from the classical picture, as studied in ditail in Sec.~\ref{energy and temperature}.
In Sec.~\ref{sec:comparison} we show that at the energy scale of our concern the quantum energy level spacings
are not sufficient to judge
the quantum chaos, so the thermal OTOC is a better indicator of the quantum chaos.
Sec.~\ref{sec:conclusion} is devoted to a summary and discussions. 
In App.~\ref{sec:truncation error} we estimate numerical errors in the evaluation of the OTOCs.


\section{Review: Classical chaos in the coupled harmonic oscillators}
\label{sec:review1}

In this section we introduce the coupled harmonic oscillator (CHO) model 
\cite{Matinyan Savvidy Ter-Arutunian Savvidy (1981a), Matinyan Savvidy Ter-Arutunian Savvidy (1981b), Savvidy (1984)}, which is the system we study throughout this paper, and make a review
of its classical properties on chaos.
In Sec.~\ref{subsec:reduction}, we define the Hamiltonian of the model
and describe the dimensional reduction. 
Then, in Sec.~\ref{subsec:classical properties} we look at
the Poincar\'e sections and the classical Lyapunov exponent, whose 
energy dependence is $E^{1/4}$ at high energy $E$ \cite{Chirikov Shepelyanski (1981)}.

\subsection{The coupled harmonic oscillator model}
\label{subsec:reduction}

The quantum mechanical model which we study is defined by the Hamiltonian \eqref{CHO},
which we call CHO Hamiltonian:  $H = p_x^2 + p_y^2 + U(x,y)$ with the potential term  
\begin{align}\label{U}
	U(x,y) = \frac{\omega^2}{4}(x^2 + y^2) + g_0 x^2 y^2  \, .
\end{align}
Here $\omega$ and $g_0$ are constant parameters.
Without losing its generality, we can take $\omega=1$ by a rescaling of the
dynamical variables $x(t)$ and $y(t)$ and the Hamiltonian. 
In this paper, we take $g_0=1/10$, for our purpose of observing the chaotic and
regular phases quantum mechanically\footnote{
If one uses a larger $g_0$ instead, the low energy regular phase may not be
seen,
because the quantum zero-point energy is too close to the classical
regular-chaos transition energy scale (see Sec.~\ref{subsec:classical properties}).
}.
This is a coupled harmonic oscillator system, and the coupling is given by
the non-linear term $g_0 x^2y^2$.

To motivate readers, we here show that this Hamiltonian \eqref{CHO}
can be obtained by a dimensional reduction of a four-dimensional 
SU(2) Yang-Mills-Higgs theory in a flat spacetime\footnote{
Historically, homogeneous classical Yang-Mills theory was solved in \cite{Baseian:1979zx},
and for early papers on the quantum treatment, see \cite{Luscher:1982ma,Savvidy:1984gi}.
}. 
The Lagrangian 
of the latter is given by
\begin{align}
\label{YMH Lagrangian}
	{\cal{L}} = (D_\mu \phi)^\dagger (D^\mu \phi) - V(\phi) - \frac{1}{4}F^a_{\mu \nu}F^{\mu \nu a}
\end{align}
where $D_\mu \equiv \partial_\mu-igA_\mu^a T^a$ ($a=1,2,3$), 
$T^a \equiv \sigma^a/2$ (with the Pauli matrices $\sigma^a$), 
$F_{\mu\nu}^a \equiv \partial_\mu A^a_\nu-\partial_\nu A^a_\mu+g\epsilon^{abc}A_\mu^b A_\nu^c$. 
The Levi-Civita symbol $\epsilon^{abc}$ is a totally antisymmetric tensor 
with $\epsilon^{123}=1$. The scalar field potential $V(\phi)$ in \eqref{YMH Lagrangian} is
\begin{align}
	V(\phi) = \mu^2 |\phi|^2 + \lambda |\phi|^4 \, .
\end{align}
Although any quantum property of the Yang-Mills theory is quite intricate,
after the dimensional reduction it becomes a tractable quantum model. 
First, we choose a gauge condition $A_0^a=0$, and assume that the fields are spatially homogeneous;
\begin{align}
	\partial_k \phi = 0 \, , \quad \partial_k A^a_l = 0 \, .
\end{align}
Here $k,l$ denote spacial indices. 
In the vacuum which is spontaneously broken with $\mu^2<0$, 
we turn on only $A_1^1\equiv x(t), A_2^2\equiv y(t)$, then the Hamiltonian
of the system reduces to
\begin{align}
	H = \frac{1}{2}(\dot{x}^2+\dot{y}^2) + \frac{g^2 v^2}{8}(x^2 + y^2) + \frac{g^2}{2}x^2 y^2 \, , 
\end{align}
where $v$ ($\neq 0$) corresponds to the vacuum expectation value 
of the original scalar field $\phi$. 
A simple rescaling brings this Hamiltonian to \eqref{CHO}.

It is known that, regardless of whether classical or quantum,
this system \eqref{CHO} is in 
a regular phase at low energy and in a chaotic phase at high energy.
This property is easily understood from the structure of the potential $U(x,y)$. 
On the $(x,y)$-plane, in the region close to the origin, the potential is 
well approximated by the first term in $U(x,y)$ in \eqref{U}.
Thus at the low energy, this system behaves as a two dimensional harmonic oscillator, and  is in the regular phase. 
On the other hand, at the high energy, the energy contribution of the non-linear term becomes significant and this system turns into chaotic.

The following classical and quantum analyses have validated these phases.
In the classical analysis \cite{Matinyan Savvidy Ter-Arutunian Savvidy (1981b), Pullen Edmonds (1981),Chirikov Shepelyanski (1981)}, the  
Poincar\'{e} sections and Lyapunov exponents, which are useful to see if a classical system is chaotic or not, were examined. We review these quantities in 
the next subsection. 

Quantum mechanically,
the distribution of nearest-neighbor spacings of the energy eigenvalues
can discriminate chaoticity of the system. 
Regular systems show a Poisson distribution, while  chaotic systems
show a Wigner one.
In the quantum analysis \cite{Haller Koppel Cederbaum (1984)}, the distribution of the energy eigenvalues
of the CHO model \eqref{CHO} was shown to be Wigner-like (Poisson-like) 
at high (low) energy.

In quantum systems, 
Out-of-Time-Order Correlator (OTOC) is another quantity to discriminate a quantum chaos. 
The quantitative indicator of quantum chaos is 
the quantum Lyapunov exponent, which measures 
the exponential growth of the OTOC. 
In this paper we focus on the OTOC of the CHO system \eqref{CHO},
and evaluate the ability of it as an indicator of the quantum chaos,
compared to the energy eigenvalue distribution method.

\begin{figure}[t]
\centering
	\includegraphics[width=125mm]{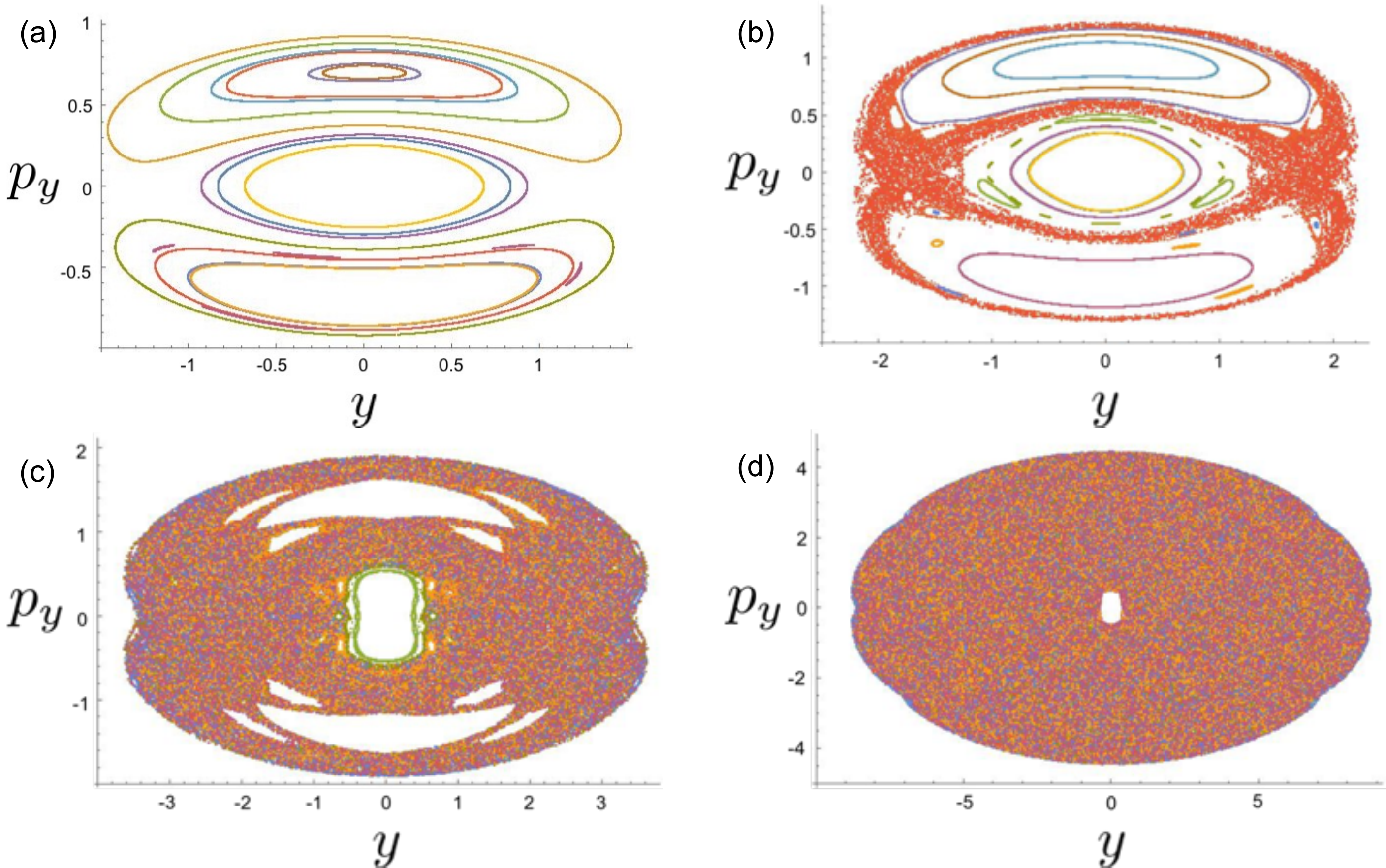}
	\caption{Poincar\'{e} sections of the CHO Hamiltonian system. (a)$E=1$: The orbits are periodic and the system is integrable. (b)$E=2$: Scattering has begun and the chaos is appearing. (c)$E=4$: Although much scattered, some periodic orbits can be seen. (d)$E=20$: Almost everywhere scattered, the system is chaotic enough. Points are sampled when the trajectory crosses $x=0$ with $p_x>0$. 
	}
	\label{fig:Poincare section}
\end{figure}


\subsection{Energetic view of Poincar\'{e} sections and classical Lyapunov exponents}
\label{subsec:classical properties}

Drawing Poincar\'{e} sections is a popular method to find qualitatively a classical chaos.
To illustrate concretely the classical chaos of the CHO Hamiltonian system \eqref{CHO},
following 
\cite{Matinyan Savvidy Ter-Arutunian Savvidy (1981b), Pullen Edmonds (1981)},
we plot in Fig.\ref{fig:Poincare section} the Poincar\'{e} sections at several chosen values of the energy.
In the figure, subfigure (a), (b), (c) and (d) correspond to $E=1, 2, 4$ and $20$ respectively.
At a low energy, we see the orbits in the Poincar\'e section, so the system is in a regular phase.
On the other hand, at high energy, 
orbits are destroyed and the Poincar\'e section is filled with scattered 
plots, meaning that the system is in a chaos phase.
When one increases the energy from zero, 
the decay of the orbits starts to show up at $E\lesssim 2$, and the whole Poincar\'e section
is almost covered by the scattered plots at $E\gtrsim 5$.
Hence the classical CHO Hamiltonian system has a transition between
the regular phase and the chaos phase. 
To be more precise, 
in our system \eqref{CHO} with the chosen parameters $\omega=1$ and $g_0=1/10$,
the transition energy scale between the regular and the mixed phases is around $E\simeq 2$,
and the one between the mixed and the complete chaos phase is around $E \simeq 10$.
In the last section, we will compare our results of the 
quantum chaotic indicator OTOC with this classical phase structure.

\begin{figure}[t]
\centering
	\includegraphics[width=75mm]{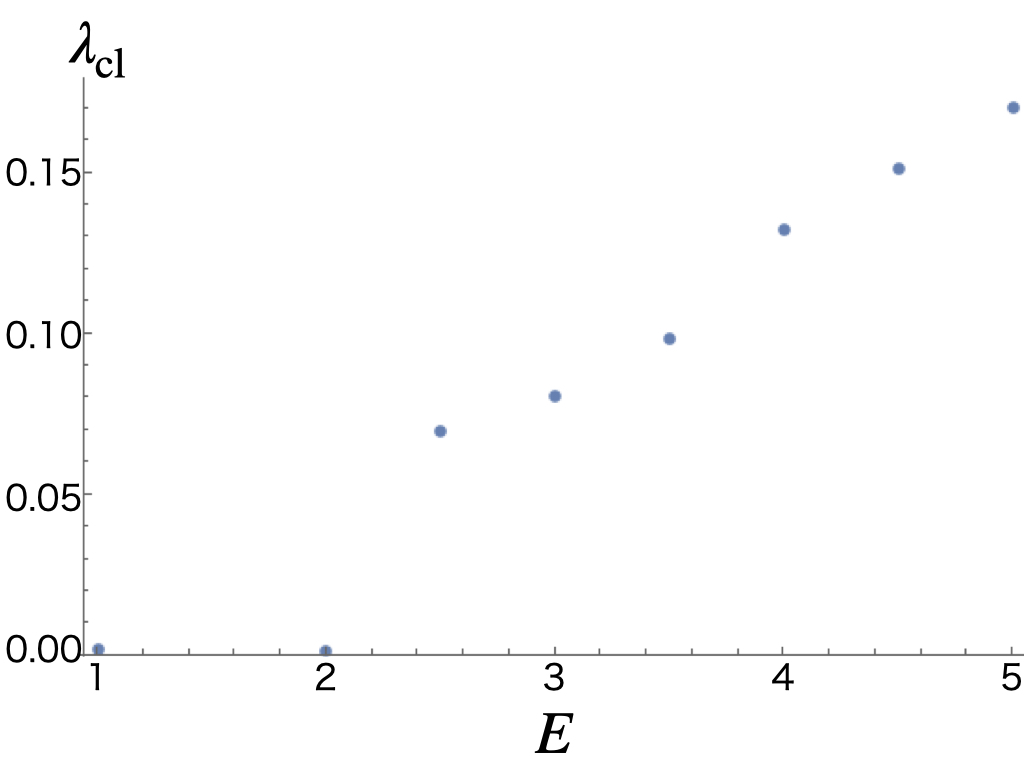}
	\caption{The classical Lyapunov exponents of our CHO Hamiltonian system \eqref{CHO} as 
	a function of the energy of the initial condition. 
	Note that we randomly chose the initial conditions for the numerical calculation of the Lyapunov exponents, 
	and did not evaluate the maximum value of them for a given energy. This figure is for
	a rough understanding of the classical behavior of chaos of our system, and for the estimation 
	of the value of the Lyapunov exponents. }
	\label{fig:classical Lyapunov}
\end{figure}

The classical Lyapunov exponent $\lambda_{\rm{cl}}$ measures how much a given classical system is chaotic.
It characterizes the initial value sensitivity of the orbits in the phase space 
as an exponential growth $\Delta x(t) \simeq \Delta x(0) e^{\lambda_{\rm{cl}}t}$,
where $\Delta x(t)$ is the distance between two nearby orbits. Any positive $\lambda_{\rm cl}$
means that the system is chaotic, and furthermore, the Lyapunov exponent can quantitatively
evaluate the chaos. 
In Fig.~\ref{fig:classical Lyapunov} we show a set of numerically calculated\footnote{We used the mathematica LCE package formulated by M. Sandri.} classical Lyapunov exponents of the CHO Hamiltonian system \eqref{CHO}.  
At a low energy we find $\lambda_{\rm{cl}}$ vanishes, which means that the system is regular. 
On the other hand, at a larger value of the energy the Lyapunov exponent is nonzero, thus chaos shows up.
The measured Lyapunov exponent is an increasing function of energy, so the chaos is stronger for larger values of the 
energy.
The energy dependence of the Lyapunov exponents is consistent with what we observed by using the Poincar\'e sections.

In fact, when the potential energy is just given by the interaction term, $U(x,y)= g_0 x^2 y^2$,
the energy dependence of the classical Lyapunov exponent can be analytically derived \cite{Chirikov Shepelyanski (1981)}:
\begin{align}\label{energy dependence}
	\lambda_{\rm{cl}}(E) \propto E^{1/4}\, .
\end{align}
Ignoring the mass term $(\omega^2/4)(x^2+y^2)$ in the potential energy $U(x,y)$ is equivalent to going to the high energy limit,
so we can expect that at high energy our system \eqref{CHO} 
approximately follows the energy dependence \eqref{energy dependence}.
The relation \eqref{energy dependence} is derived  as follows.
Ignoring the mass term in the Hamiltonian (\ref{CHO}), we have the energy
\begin{align}
\label{E}
	E = p_x^2 + p_y^2 + g_0 x^2 y^2 \, .
\end{align}
This system allows the following scaling transformation\footnote{
The scaling law is valid even for the whole Yang-Mills theories and their supersymmetric generalizations. 
In particular, the scaling law for the D0-brane matrix theory \cite{Banks:1996vh} was found in \cite{Jevicki:1998yr, Jevicki:1998ub} and contributed to the early development in the study of M-theory.
}:
\begin{align}
(x,y) \to (\lambda x, \lambda y), \quad
E \to \lambda^4 E \, , \quad t \to \lambda^{-1} t \, ,
\label{betascale}
\end{align}
with an arbitrary positive constant $\lambda$. Since the Lyapunov exponent has the dimension of inverse time,
from the last two relations in \eqref{betascale}, it is clear that it should scale as \eqref{energy dependence} as a function of energy.

\section{Review: numerical computation of OTOC}
\label{sec:review2}

In this section we summarize methods to numerically calculate OTOCs, based on \cite{Hashimoto:2017oit}. 
First in Sec.~\ref{subsec:OTOC},  we define the OTOC which we focus on throughout this paper, and we verify that analytic expression of the OTOC for a harmonic oscillator is available, as its Heisenberg equation is exactly solvable. In general systems, however, it is difficult to solve the Heisenberg equation analytically, so we have to rely on a numerical method. We review the numerical methods for evaluating the OTOCs in generic quantum mechanical systems in Sec.~\ref{subsec:numerics}.

\subsection{Microcanonical OTOC and thermal OTOC}
\label{subsec:OTOC}

Let us take a quantum system which is described by a time-independent Hamiltonian $H = H(x_1, \cdots, x_N, p_1, \cdots, p_N)$,  in the Heisenberg picture. In the following we write $x \equiv x_1, p \equiv p_1$, and denote the $x$ operator in the Heisenberg representation at time $t$ as $x(t) \equiv e^{iHt} \, x \, e^{-iHt}$.  
The OTOC of our concern is defined as
\begin{align}
\label{OTOC}
	C_T(t) \equiv - \langle [x(t), p]^2 \rangle_T \, .
\end{align}
Here $\langle \cdots \rangle$ denotes the thermal average with the inverse temperature $\beta\equiv 1/T$,
\begin{align}
	\langle {{O}} \rangle_T \equiv \frac{1}{Z(T)} {\rm{Tr}}[e^{-\beta H}{{O}}] \, , \quad Z(T) \equiv {\rm{Tr}}[e^{-\beta H}] \, .
\end{align}
Assuming discrete energy eigenvalues, we have the following expression by using energy eigenvectors,
\begin{align}
\label{thermal and microcanonical}
	C_T(t) = \frac{1}{Z(T)}\sum_{n} e^{-\beta E_n}c_n(t) \, , 
	\quad c_n(t) = -\langle n | [x(t), p]^2 | n \rangle \, ,
\end{align}
where $H|n\rangle = E_n |n\rangle$ with a non-negative integer $n$. We call $C_T(t)$ a ``thermal OTOC'' and $c_n(t)$ a ``microcanonical OTOC.'' In a quantum chaotic system, it is expected
that the thermal OTOC grows exponentially as $C_T(t) \simeq e^{2\lambda_{\rm{q}}(T)t}$, where  $\lambda_{\rm{q}}(T)$, which we call ``quantum Lyapunov exponent,'' is positive. This exponent is a quantum analogue of the classical Lyapunov exponent described in the previous section.

As a concrete example, we consider a harmonic oscillator in one dimension,
\begin{align}
	H = p^2 + \frac{\omega^2}{4}x^2 \, .
\end{align}
In this case it is possible to solve the Heisenberg equation analytically, and the solution is
\begin{align}
\label{solution}
	x(t) = x \cos{\omega t} + \frac{2}{\omega} p \sin{\omega t}, \quad p(t) = p \cos{\omega t} - \frac{\omega}{2} x \sin{\omega t} \, .
\end{align}
Substituting these solutions and using the canonical commutation relation, we get
\begin{align}
\label{OTOC of HO}
	c_n(t) = \cos^2{\omega t}, \quad C_T(t) = \cos^2{\omega t} \, .
\end{align}
A higher dimensional harmonic oscillator has the same expression since the degrees of freedom in the other dimensions are decoupled from the sector of $(x_1, p_1)$.
The OTOCs of the harmonic oscillator are oscillatory, thus do not show any exponential growth. This is consistent with the fact that this system is solvable and regular. In addition, they depend on neither the mode label $n$ nor the temperature $T$.

\subsection{Numerical method for OTOC}
\label{subsec:numerics}

In general quantum systems it is difficult to solve their Heisenberg equation, so we rely on numerical methods to evaluate the OTOCs. Here,
we rewrite the OTOCs \eqref{thermal and microcanonical} for our later purposes to performing the numerical calculations.

The completeness condition $\sum_n |n\rangle \langle n |=1$ brings the microcanonical OTOC $c_n(t)$ in \eqref{thermal and microcanonical} to the following form:
\begin{align}
\label{microcanonical}
	c_n(t) = \sum_m b_{nm}(t)b_{nm}^*(t) \, , \quad b_{nm}(t) \equiv -i\langle n | [x(t),p] | m \rangle \, .
\end{align}
Substituting $x(t)=e^{iHt} x e^{-iHt}$ to this expression of $b_{nm}(t)$ and using the completeness condition again, we obtain
\begin{align}
	b_{nm}(t) = -\sum_k (e^{iE_{nk}t}x_{nk}p_{km}-e^{iE_{km}t}p_{nk}x_{km}) \, .
\end{align}
Here we have defined $x_{nm} \equiv \langle n | x | m \rangle$, $p_{nm} \equiv \langle n | p | m \rangle$, and $E_{nm} \equiv E_n - E_m$.
When the Hamiltonian is of the form (which includes the case of the CHO Hamiltonian \eqref{CHO})
\begin{align}
	H = \sum_{i=1}^N p_i^2 + U(x_1,\cdots,x_N) \, ,
\end{align}
one can easily verify $[H,x]=-2ip$. Then evaluating the matrix elements of this relation with $\langle n |$ and $| m \rangle$, we find $p_{nm} = \frac{i}{2}E_{nm}x_{nm}$. By using this, we derive a useful expression
\begin{align}
\label{b}
	b_{nm}(t) = \frac{1}{2}\sum_k x_{nk}x_{km}(E_{km}e^{iE_{nk}t}-E_{nk}e^{iE_{km}t}) \, .
\end{align}
With these, OTOCs can be numerically evaluated, following the concrete steps:
\begin{itemize}
\item[1.]  Solve the Schr\"{o}dinger equation of the given system to obtain the energy eigenvalues and the wave functions.
\item[2.]  Compute $x_{nm} = \langle n | x | m \rangle$ with numerical integration.
\item[3.]  Substitute the result of 2 to \eqref{b} to calculate $b_{nm}(t)$.
\item[4.]  Evaluate the microcanonical OTOC $c_n(t)$ by substituting the result of 3 to \eqref{microcanonical}.
\item[5.]  Evaluate the thermal OTOC $C_T(t)$ by using \eqref{thermal and microcanonical}.
\end{itemize}
In these numerical evaluations, approximations by introducing a finite cut-off to the infinite sums are necessary\footnote{We check that these truncations do not affect our results, see Appendix \ref{sec:truncation error}.}.
In the next section, we shall use this strategy to numerically evaluate the OTOCs in the CHO system.


\section{Exponential growth of thermal OTOC }
\label{sec:exponential growth of OTOC}

In this section, we numerically evaluate the OTOCs of the CHO system following the method given in  Sec.~\ref{subsec:numerics}, and find that the growth of the thermal OTOC in early time is exponential.
We read the quantum Lyapunov exponent at each temperature data, and study its temperature dependence.

\begin{figure}[t]
\centering
	\includegraphics[width=90mm]{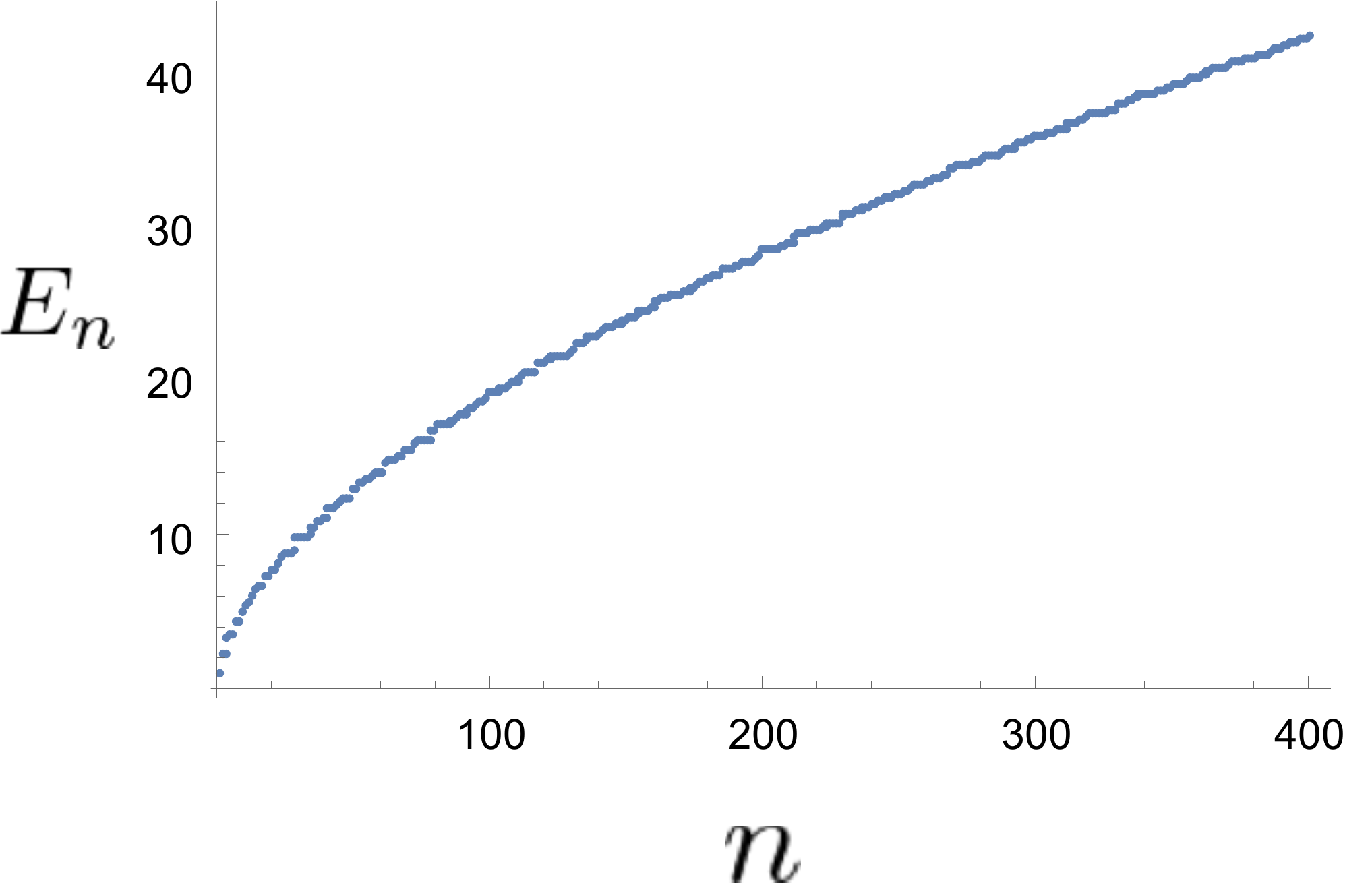}
	\caption{Energy eigenvalues of the CHO Hamiltonian with $\omega=1$ and $g_0=1/10$.
	}
	\label{energy_level}
\end{figure}

\subsection{Preparations: microcanonical OTOC}
As a preparation for evaluating the thermal OTOC from the next subsection, here we describe our results of the microcanonical OTOC obtained through the
numerical strategy 1,2,3 and 4 given in the previous section.
First, we solve the time-independent Schr\"{o}dinger equation\footnote{For solving it, we use Mathematica package NDEigensystem.}
\begin{align}
	-\left(\frac{\partial^2}{\partial x^2}+\frac{\partial^2}{\partial y^2}\right) \psi_n(x, y) + \left[ \frac{\omega^2}{4} (x^2+y^2) + g_0 x^2 y^2 \right] \psi_n(x, y) = E_n \psi_n(x, y) \, ,
	\label{Schr}
\end{align}
and obtain the energy eigenvalues $E_n$ and the wave functions $\psi_n(x,y)$. See Fig.~\ref{energy_level} for the obtained distribution of the energy levels.

\begin{figure}[t]
\centering
	\includegraphics[width=100mm]{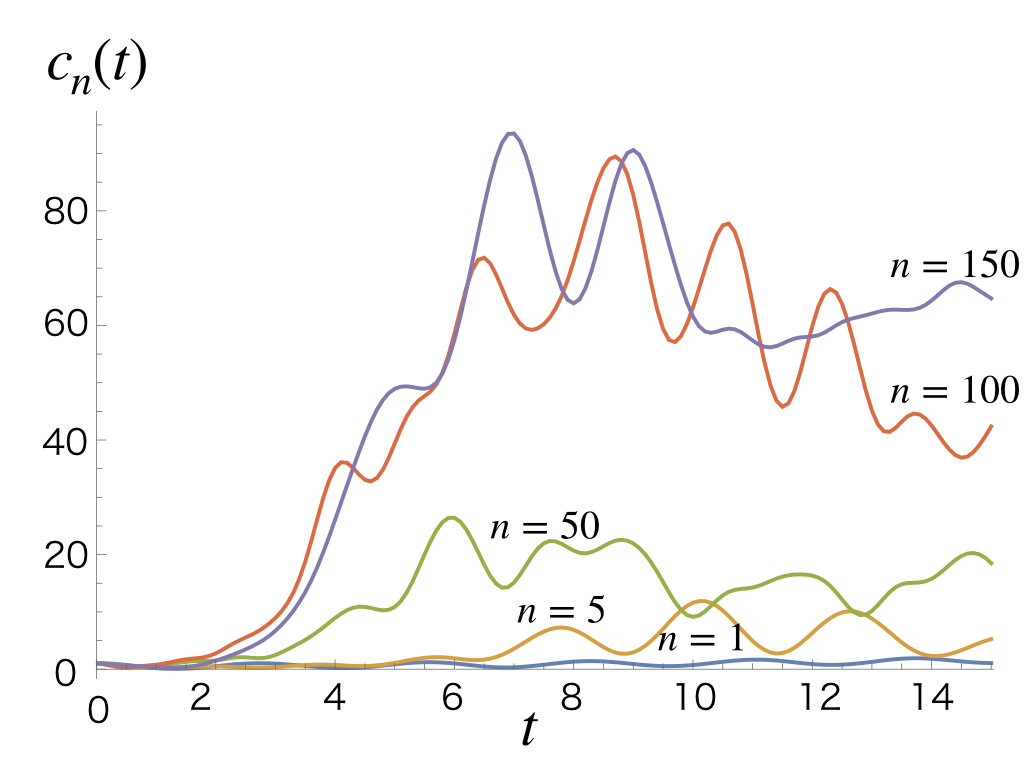}
	\caption{Numerical results of the microcanonical OTOC of the CHO Hamiltonian system for the energy level $n=1,5,50,100,150$. Low modes behave periodic in time, while higher modes deviate from the periodic behavior.}
	\label{fig:microcanonical OTOC}
\end{figure}

By using them, we evaluate the microcanonical OTOC. Fig.~\ref{fig:microcanonical OTOC} is our numerical result.  The truncation errors in this evaluation will be dealt with in Sec.~\ref{sec:truncation error}.

The resultant microcanonical OTOC for low modes oscillates almost periodically in time, and any exponential growth of it cannot be found. This behavior is similar to the microcanonical OTOC of a harmonic oscillator \eqref{OTOC of HO}. 
On the other hand, for higher modes, the behavior of the microcanonical OTOC deviates from that of the harmonic oscillator. This is expected from the structure of the potential $U(x,y)$ of \eqref{CHO}. The potential $U(x,y)$ is well approximated by a harmonic oscillator around the origin of the $(x,y)$ plane, so the wave functions localized in the vicinity of the origin behave like that of the harmonic oscillator, which happens at low energy. On the contrary, for the higher modes, the the wave functions spread, and the second term of the potential $U(x,y)$ contributes dominantly. This explains the 
deviation of the microcanonical OTOC from the harmonic oscillator  (\ref{OTOC of HO}).


\subsection{Reading quantum Lyapunov exponent from thermal OTOC}

By substituting the numerical results of the microcanonical OTOC $c_n(t)$ to the expression \eqref{thermal and microcanonical}, we compute the thermal OTOC $C_T(t)$. The numerical results are shown in Fig.~\ref{fig:thermal OTOC}. The obtained thermal OTOC shows two apparent characteristics: at the early time stage it grows, and at the late time stage it saturates to a constant value. In Sec.~\ref{late time behavior} we study that the latter part reproduces the expected property in any quantum chaotic system. Then in Sec.~\ref{quantum Lyapunov} we show that the growth of the thermal OTOC at early time is  exponential.


\subsubsection{Reproducing late-time behavior of OTOC}
\label{late time behavior}

It is known that in generic quantum chaotic systems, the correlation between $x(t)$ and $p$ is gradually 
lost in the time evolution due to the chaoticity. 
After the Ehrenfest time, the value of the thermal OTOC \eqref{OTOC} is expected to approach 
asymptotically in time \cite{Maldacena:2015waa}
\begin{align}
\label{asymptotic value}
	C_T(\infty) = 2\langle x^2 \rangle_T \langle p^2 \rangle_T \, .
\end{align}
We shall use this first for checking our numerical evaluation of the thermal OTOC. 

\begin{figure}[t]
	\begin{minipage}{0.48\hsize}
	\centering
		\includegraphics[width=70mm]{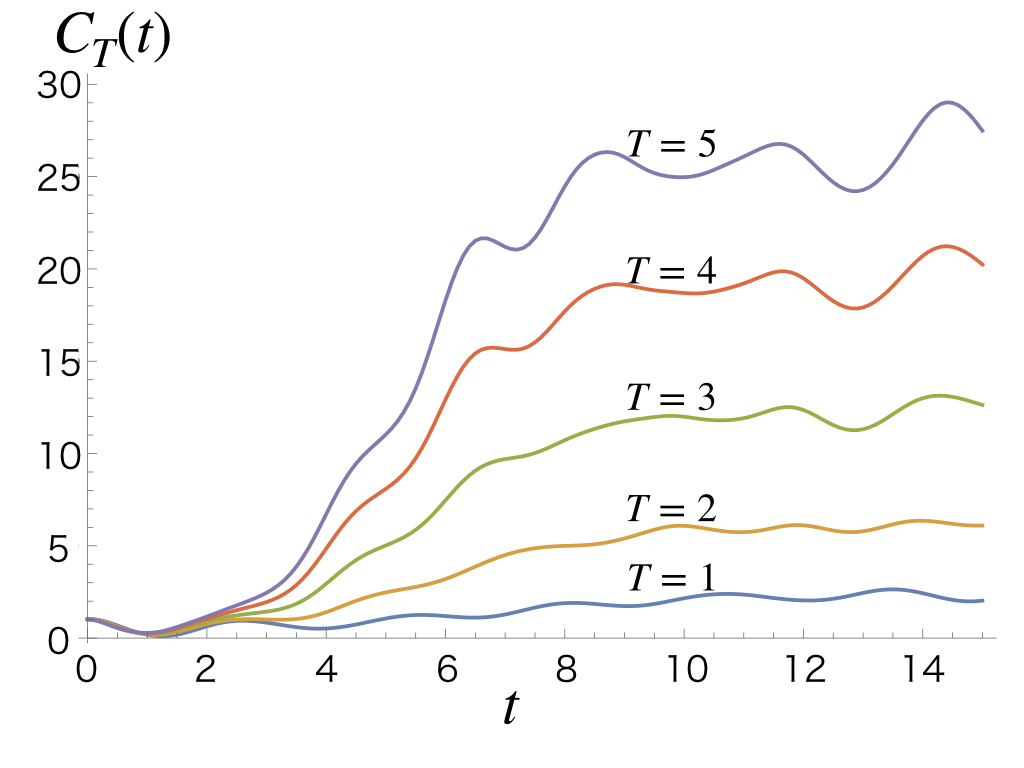}
		\caption{Time evolution of the thermal OTOC for various values of the temperature $T$.}
		\label{fig:thermal OTOC}
	\end{minipage}
	\hspace{3mm}
	\begin{minipage}{0.48\hsize}
	\centering
		\includegraphics[width=70mm]{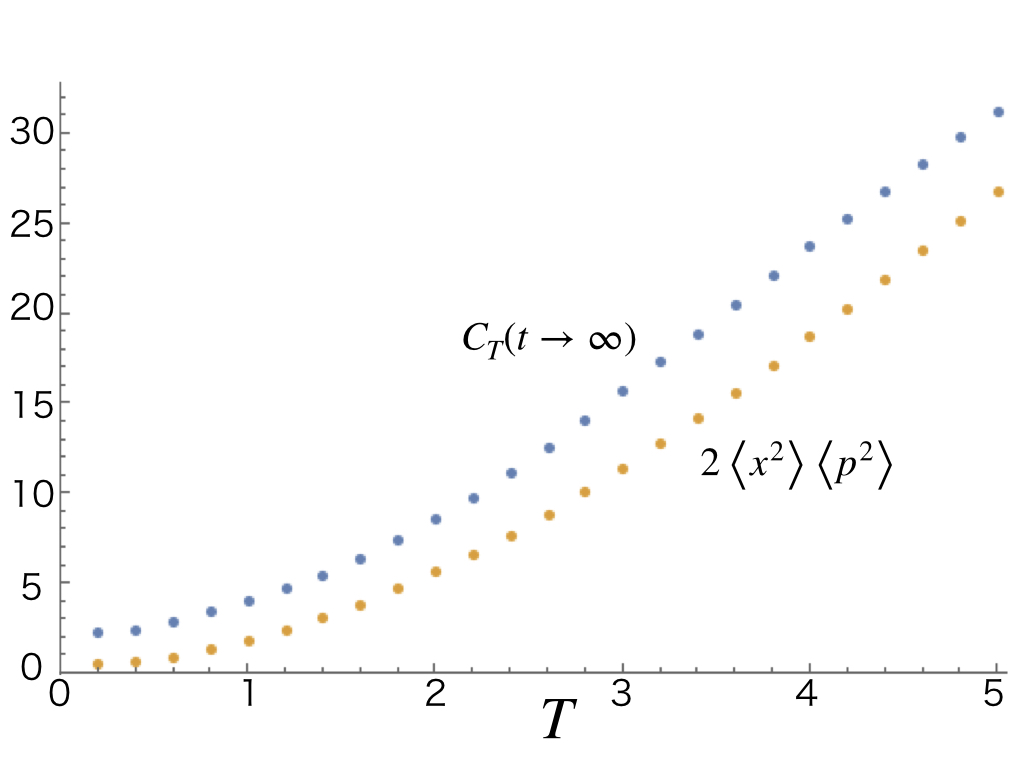}
		\caption{Asymptotic values of the numerical results of the thermal OTOC and the property of the OTOC in a quantum chaotic system \eqref{asymptotic value}. }
		\label{fig:asv}
	\end{minipage}
\end{figure}

Let us look at  our numerical results of the thermal OTOC, Fig.~\ref{fig:thermal OTOC}. 
At low temperature, the thermal OTOC does not grow so much, as in the case of the low mode microcanonical OTOC. 
This is because, in \eqref{thermal and microcanonical} at low temperature, the contribution from high modes is suppressed due to the Boltzmann factor $e^{-\beta E_n}$. 
On the other hand, at higher temperature, the Boltzmann weight does not suppress the high modes, then more number of modes contribute to the thermal OTOC.
Indeed, our numerical results in Fig.~\ref{fig:thermal OTOC} show that at late times the time evolution of the value of the OTOCs is flattened, and look like they approach some constants. 

To find out quantitatively the asymptotic values, 
we evaluate the value of the OTOC for each temperature by averaging its values at $t=20,22,24,26,28,30$. 
In Fig.~\ref{fig:asv} we plot those asymptotic values evaluated from our numerical results of the thermal OTOC, as a function of the temperature. For a comparison, we also plot the temperature dependence of the expected expression \eqref{asymptotic value}. Their trends are quite similar to each other. Thus we conclude that the saturated parts of the thermal OTOC indicate that the CHO Hamiltonian system is quantum chaotic in the temperature region we consider\footnote{The slight difference from
\eqref{asymptotic value} could come from our truncation of energy levels, and also from the finite time effect.}.

\subsubsection{Very early stage of OTOC}
\label{sec:Very early}

Next we focus on the early time part of the thermal OTOC in Fig.~\ref{fig:thermal OTOC}. 
At the very early stage, the thermal OTOC decreases, and the temperature dependence is not observed. Then after a while the temperature dependence shows up. To extract the quantum Lyapunov exponent due to the quantum chaos later, we shall reveal the physical origin of this behavior at the very early stage.

At sufficiently early times, the contribution of the nonlinear term $\frac{1}{10}x^2 y^2$ of CHO Hamiltonian \eqref{CHO} can be treated as a perturbation;
\begin{align}
	H = H_0 + g_0x^2 y^2, \quad H_0 = p_x^2 + p_y^2 + \frac{\omega^2}{4}(x^2+y^2) \, ,
\end{align}
with $g_0=1/10$ and $\omega =1$.
For that reason, the very early time behavior should be identical to that of the harmonic oscillator. 
There exists a time scale at which the nonlinear term starts to contribute significantly. This time scale can be estimated in the following way.
For small $t$, the Heisenberg operator $x(t) = e^{iHt} x e^{-iHt}$ in the OTOC \eqref{OTOC} can be approximated by $x(t) \simeq (1 + ig_0x^2 y^2t)x_0(t)(1 - ig_0x^2 y^2t)$.
Here $x_0(t)$ is the Heisenberg operator evolved by the free Hamiltonian $H_0$, which is given in \eqref{solution}. We adopt a principle that when the expectation value of $g_0 x^2 y^2t$ for the ground state $| 0 \rangle$ of $H_0$ grows roughly to ${\cal O}(10\%)$, the perturbation expansion given above is broken. Evaluating this with the harmonic oscillator ground state, we find
\begin{align}
	\langle 0 | g_0 x^2 y^2t | 0 \rangle = g_0 t \, ,
\end{align}
therefore we expect that, since our coupling constant is $g_0=1/10$, $t\simeq1$ is the time when the nonlinear term starts to contribute to the value of the thermal OTOC. This estimation is consistent with the numerical results shown in Fig.~\ref{fig:thermal OTOC}.


\subsubsection{Reading quantum Lyapunov exponent}
\label{quantum Lyapunov}

\begin{figure}[t]
\centering
	\includegraphics[width=90mm]{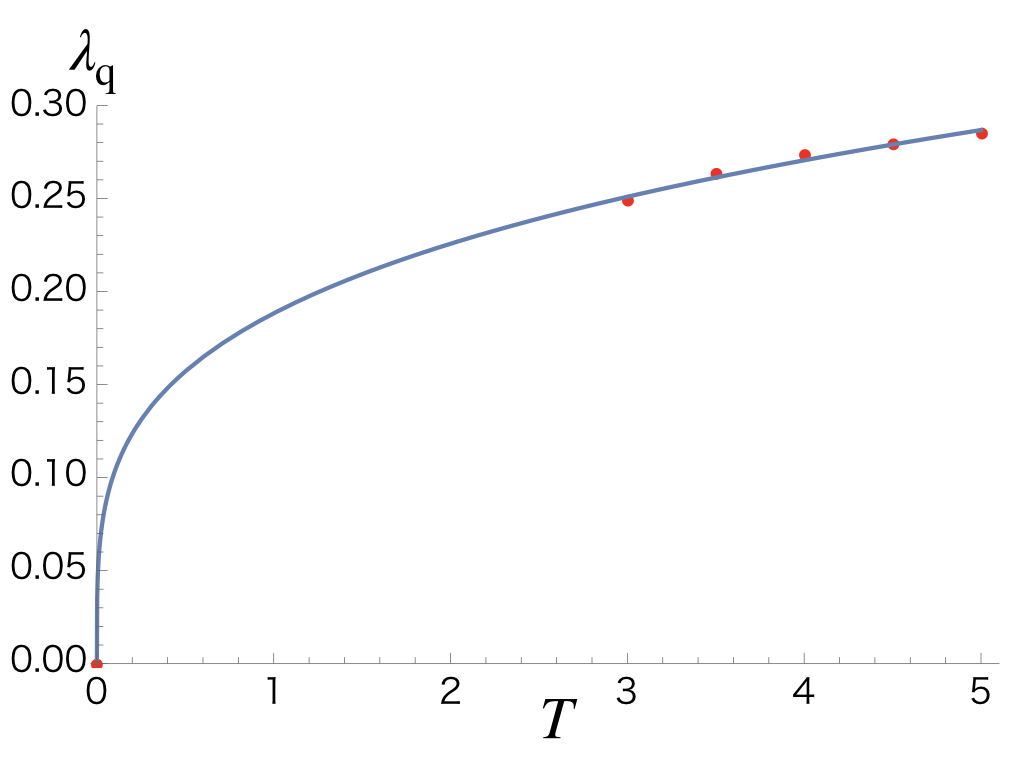}
	\caption{Measured quantum Lyapunov exponent as a function of temperature, and its fitting curve. 
	}
	\label{fig:quantum Lyapunov}
\end{figure}

We are ready to read the quantum Lyapunov exponent from the thermal OTOC. 
Our numerical results in Fig.~\ref{fig:thermal OTOC} show that the time evolution behavior of the OTOC is complicated, but we can use the observations in Sec.~\ref{late time behavior} and Sec.~\ref{sec:Very early} to extract the time regions for which the 
OTOC is expected to evolve exponentially in time.
As seen in Sec.~\ref{subsec:classical properties}, the classical CHO system is subject to a phase transition from the regular to the chaotic phase,
when going from lower to higher energy scales.
Equating the energy roughly to the temperature, the ``classical phase transition temperature'' is about $T\simeq 3$. So, it is expected that the thermal OTOC can grow exponentially for  $T \geq 3$. Furthermore, in Sec.~\ref{late time behavior} and Sec.~\ref{sec:Very early}, we observe that for $t\leq 2$ the nonlinear effect is not strong enough, while for $t\geq 6$ (which is the expected Ehrenfest time) the system saturates to the asymptotic value of the OTOC. 
From these observations, we make a numerical fitting of the data of our thermal OTOC for $T\geq 3$ in the time domain\footnote{
To judge whether the evolution is linear or exponential, we need the time period 
which is not much shorter than the inverse of the Lyapunov exponent.
Our time period is $\delta t = 6-2=4$, while we will see 
that our quantum Lyapunov exponent of the thermal OTOC gives $1/\lambda \sim 4$,
which is almost equal to the time period.
Also, note that for the case of the quantum stadium billiard \cite{Hashimoto:2017oit}, the growth of the thermal OTOC was not judged to be exponential,
because even when one takes a deformation to the regular limit (by bringing the billiard to a circle), the growth remains.
In our case, if one takes the deformation to the regular limit by bringing $g_0 \to 0$, the system reduces to the decoupled harmonic oscillators
so our growth in the thermal OTOC disappears. From this observation, it is plausible that the growth in our thermal OTOC is exponential and is
due to the chaos. We would like to thank K.~Murata for his valuable comment on this.
} 
$2\leq t \leq 6$, by an exponential function   $a(T) e^{2\lambda_{\rm q}(T) t}$ with the temperature-dependent quantum Lyapunov exponent $\lambda_{\rm q}(T)$.
We plot our numerical result for the quantum Lyapunov exponent for various values of the temperature, $T=3,3.5,4,4.5,5$, in Fig.~\ref{fig:quantum Lyapunov}.

Due to the following two observations about the exponents, we claim that the thermal OTOCs of the CHO system grow exponentially.

\begin{description}
\item[(1)] 
The order of magnitude of $\lambda_{\rm q}$ equals that of the classical $\lambda_{\rm{cl}}$.

From the first place, the OTOC $C_T(t)$ is defined as a quantum analogue of the classical Poisson bracket $\{ x(t), p \}^2 \simeq e^{2\lambda_{\rm{cl}}t}$,
therefore the behavior $C_T(t) \simeq e^{2\lambda_{\rm{cl}}t}$ is expected, as long as the system is still close to the classical time evolution, until Ehrenfest time.
Thus it is also natural to expect that the order of quantum Lyapunov exponent $\lambda_{\rm{q}}$ and of classical Lyapunov exponent $\lambda_{\rm{cl}}$ is same. 
Our classical Lyapunov exponent are shown in Fig.~\ref{fig:classical Lyapunov} and the values\footnote{Note that the plots in Fig.~\ref{fig:classical Lyapunov} are
just for some fixed initial conditions, and we haven't searched the largest Lyapunov exponent. So the maximal values should be a little bit larger than the values in Fig.~\ref{fig:classical Lyapunov}. Furthermore, the quantum Lyapunov exponent is for the thermal average, which includes not only the mode at $E=T$ but also the other energy modes. Thus equating these two exponents need a caution. Here we just check whether the order of magnitude for these two exponents is shared commonly or not.} are ${\cal O}(0.1) \sim {\cal O}(0.2)$, which is consistent with
the quantum values in Fig.~\ref{fig:quantum Lyapunov}.

\item[(2)] The temperature dependence of $\lambda_{\rm q}$ is similar to the energy dependence of $\lambda_{\rm{cl}}$.

We postulate that the temperature dependence of the quantum Lyapunov exponent $\lambda_{\rm q}$ is $\lambda_{\rm q}(T) = b T^c$ with some constants $b$ and $c$, and fit our numerical results with this function, see Fig.~\ref{fig:quantum Lyapunov}. The fitting function satisfies $\lambda_{\rm q}(T=0) = 0$, 
since at zero temperature there should not be the exponential growth\footnote{
By the way, at $T \to 0$, only ground state of the microcanonical OTOC contributes the thermal OTOC because of Boltzmann factor $e^{-\beta E_n}$. The wavefunction of ground state is localized in the region close to the origin on $(x,y)$ plane, and is almost the same as a harmonic oscillator's one. Therefore
\begin{align}
	C_T(t) \xrightarrow{T \to 0} c_1(t) \simeq \cos^2{t} \, ,
\end{align}
and the thermal OTOC is not expected to grow exponentially. It is natural to assume $\lambda=0$ at $T=0$. 
}.
As a result, we obtain\footnote{The error is a systematic error: the exponent $c$ depends on how many points we adopt when we fit them.}
\begin{align}
\label{temperature dependence}
	\lambda_{\rm{q}}(T) \propto T^c, \quad c=0.26\sim0.31 \, .
\end{align}
On the other hand, as shown in \eqref{energy dependence} in Sec.~\ref{subsec:classical properties}, the energy dependence of the classical Lyapunov exponent is given as $\lambda_{\rm cl} \propto E^{1/4}$ at sufficiently high energy. Regarding the typical energy at finite temperature is $E \simeq T$, we expect that the quantum Lyapunov exponent $\lambda_{\rm{q}}(T)$ is approximately proportional to $T^{1/4}$. Our result \eqref{temperature dependence} is consistent with that\footnote{The assumption $E \simeq T$ is further verified in details, in Sec.~\ref{energy and temperature}.}. 
\end{description}

The coincidence with the classical behavior is remarkable. Although in the CHO system which has only two degrees of freedom the Ehrenfest time is
expected to be rather short, we find a time domain in which the classical exponential growth can be detected in thermal OTOC. In other words,
when the classical behavior such as the energy dependence of the Lyapunov exponent is known, one can identify where the quantum thermal OTOC
behaves classically, and estimate the Ehrenfest time.

One of our major results is that the temperature dependence of the quantum Lyapunov exponent of the CHO system is found as \eqref{temperature dependence}. The temperature dependence is important as it indicates how close the system is to a black hole system \cite{Maldacena:2015waa}. 
For  the exponent of generic OTOCs\footnote{The definition of the OTOC here is different from that of \cite{Maldacena:2015waa}  as for the ordering of the Boltzmann weight operators. Here we naively compare the bound with ours, by assuming that the bound is valid also for our OTOC of the CHO system \eqref{OTOC}.} there exists a bound $2\tilde{\lambda}_{\rm q}(T) \leq 2 \pi T$ in the large $N$ limit. 
The measured quantum Lyapunov exponent $\lambda_{\rm{q}}(T)$ in our system is found to be consistent with this bound.

\begin{figure}[t]
	\centering
		\includegraphics[width=75mm]{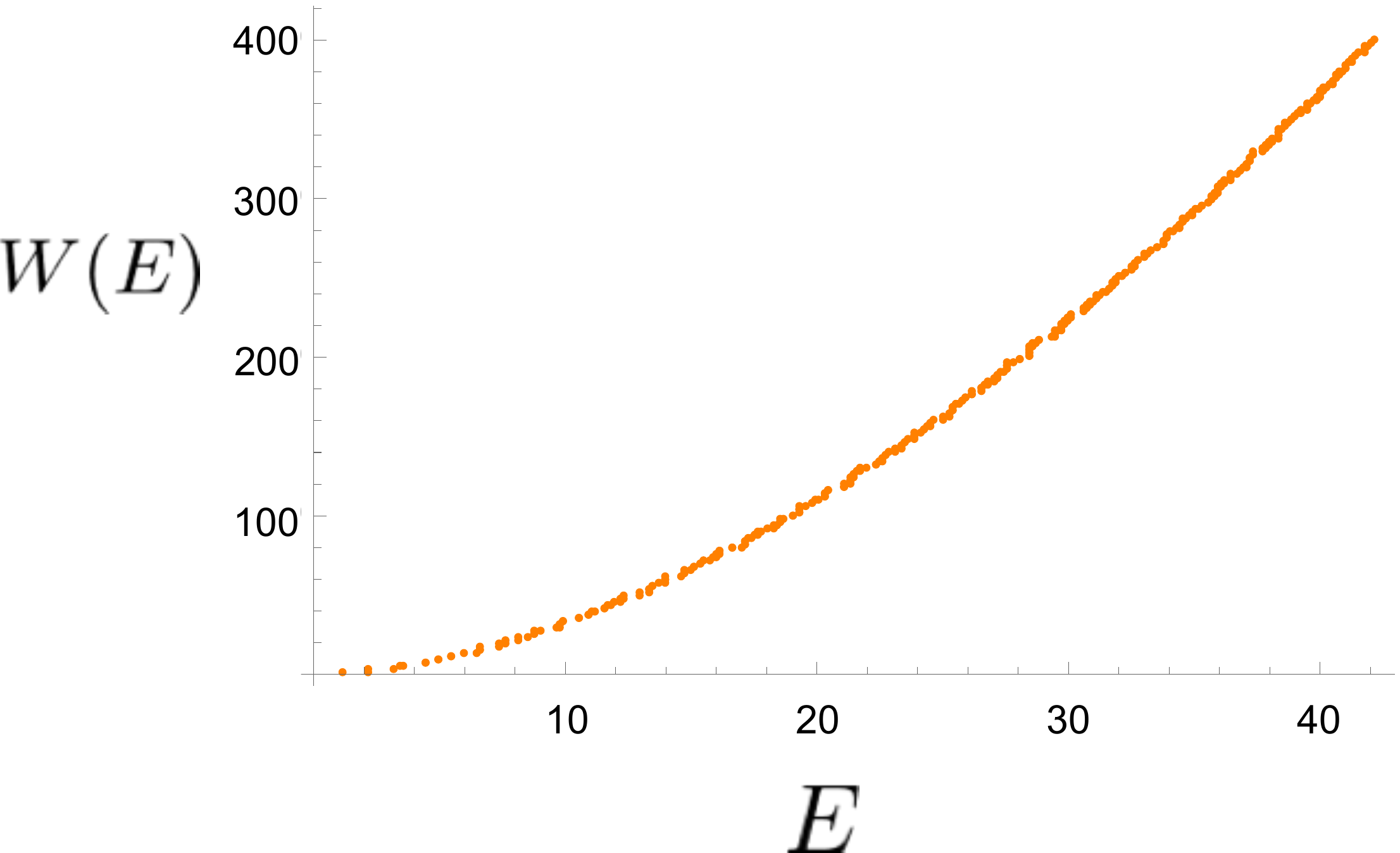}
		\includegraphics[width=75mm]{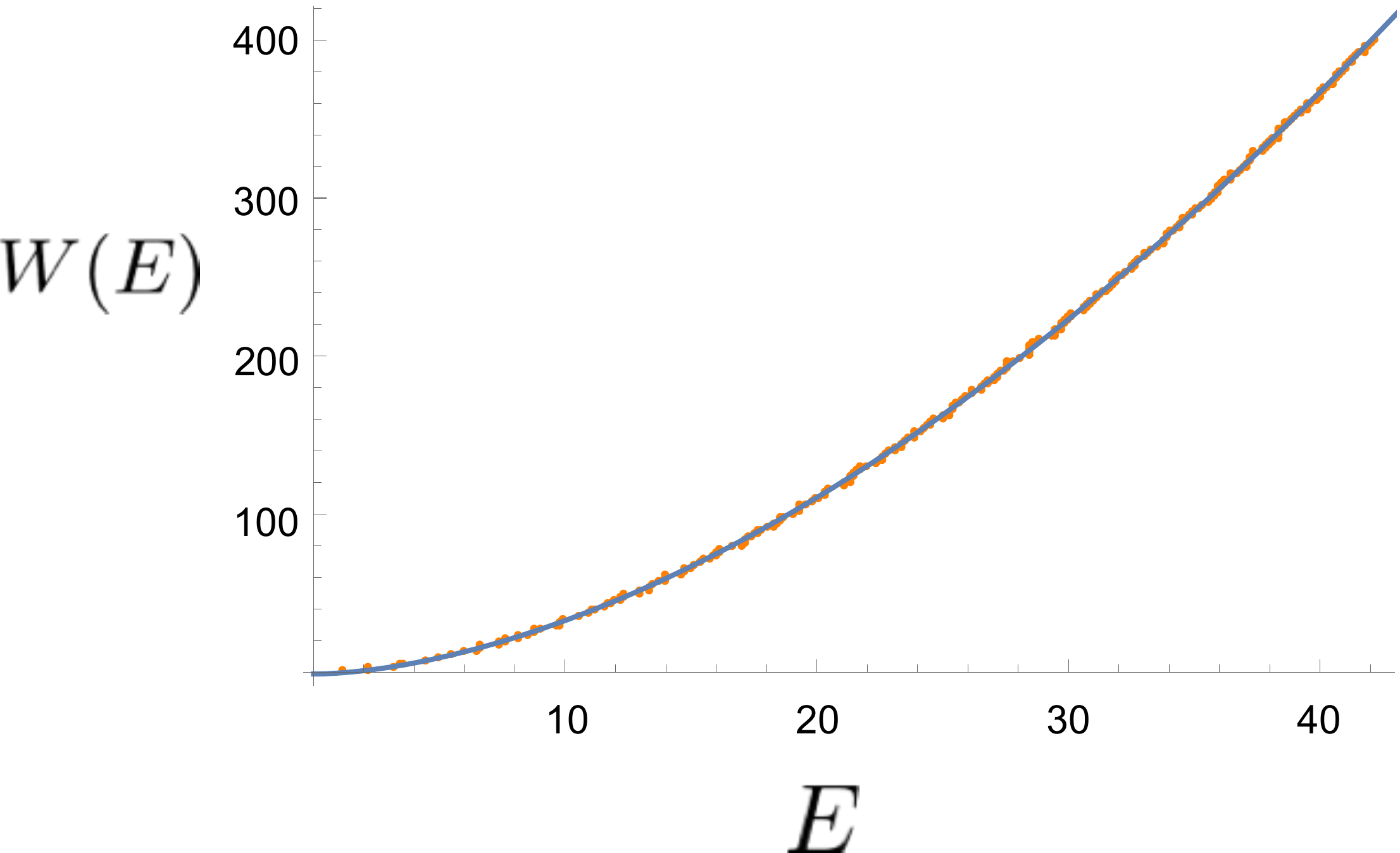}
		\caption{Left: The number of states below the energy $E$, obtained by the eigenvalues of the quantum CHO Hamiltonian system. Right: The fitting function $W(E) = -1.08 + 0.64E^{1.72}$ overlaid on the point plots of the figure Left.}
		\label{fig:the number of states}
\end{figure}


\section{Temperature and energy dependence of Lyapunov exponent}
\label{energy and temperature}

In the identification of the quantum Lyapunov exponent in Sec.~\ref{quantum Lyapunov}, we used an argument that the temperature dependence of the quantum Lyapunov exponent $\lambda_{\rm{q}}(T)$ is expected to be $\lambda_{\rm{q}}(T) \propto T^{1/4}$ due to the energy dependence \eqref{energy dependence} of the classical Lyapunov exponent $\lambda_{\rm{cl}}(E)$. In fact, this could be a too rough assumption --- as discussed in \cite{Rozenbaum:2019kdl}, the thermal average can wash off quantum chaotic information encoded in the microcanonical OTOCs. Here in this section, we elaborate on this concern in detail. We evaluate the thermal average and show by a numerical calculation that the assumption we made in Sec.~\ref{quantum Lyapunov} is reasonable.

Since the issue is the relation between the microcanonical OTOC and the thermal OTOC, and that between classical and quantum correlations, first 
we rewrite the thermal OTOC \eqref{thermal and microcanonical} as
\begin{align}
\label{OTOC in integral form}
	C_T(t) = \frac{1}{Z(T)} \int dE \rho(E)c_E(t)e^{-\beta E}\, .
\end{align}
Here, we have defined
\begin{align}
	\rho(E) = \sum_n \delta(E-E_n), \quad c_E(t) = \frac{1}{N(E)} \sum_{\substack{n \\ E_n = E}} c_n(t) \, , 
\end{align}
where $N(E)$ counts the degeneracy of the states at energy $E$. We set $c_E(t)=0$ if there exists no energy level at $E$. Next, we approximate the discrete energy level distribution of the CHO Hamiltonian system by a continuous function. We assume that $c_E(t)$ corresponds to the classical Poisson bracket $\{ x(t), p \}_P^2$, which means 
\begin{align}
\label{c_E}
	c_E(t) = e^{2\lambda_{\rm{cl}}(E)t}\, .
\end{align}
In Sec.\ref{subsec:classical properties} we have seen that at high energy the classical Lyapunov exponent $\lambda_{\rm{cl}}(E)$ shows the energy dependence \eqref{energy dependence}.
We use it here as a simple assumption that the relation \eqref{energy dependence} holds for all energy scales of our concern, and set
\begin{align}
\label{classical Lyapunov}
	\lambda_{\rm{cl}}(E) = a E^{1/4}\, .
\end{align}

\begin{figure}[t]
	\begin{minipage}{0.5\hsize}
	\centering
		\includegraphics[width=75mm]{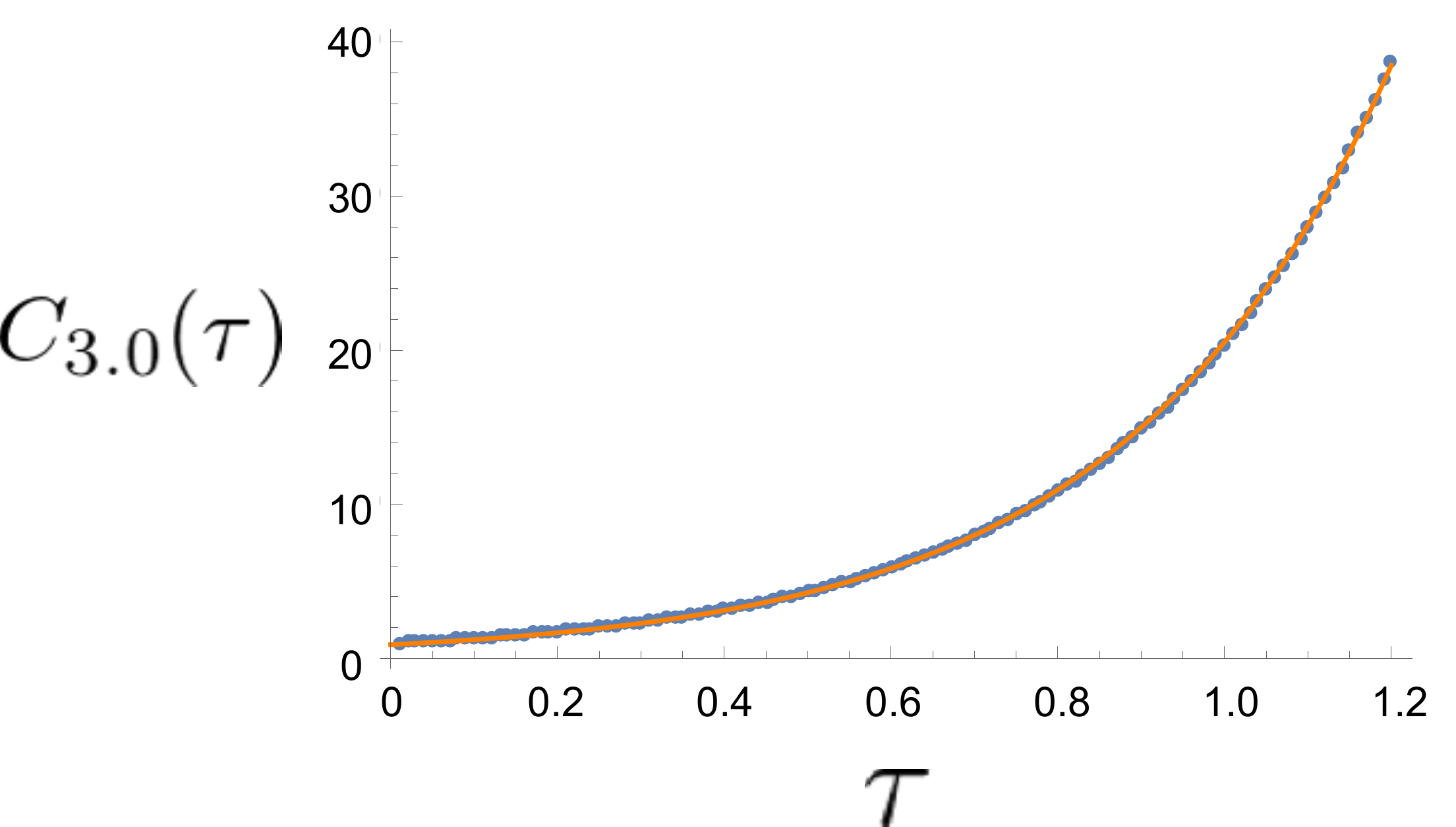}
		\caption{Points: Numerically calculated $C_T(\tau)$ in \eqref{OTOC in integral form 2}, with $T=3.0$ as a function of $\tau$. Line: a fitting of the points by $\alpha e^{L \tau}$ with constant $\alpha$ and $L$.		}
		\label{fig:exp_fit}
	\end{minipage} \hspace{3mm}
	\begin{minipage}{0.5\hsize}
	\centering
		\includegraphics[width=67mm]{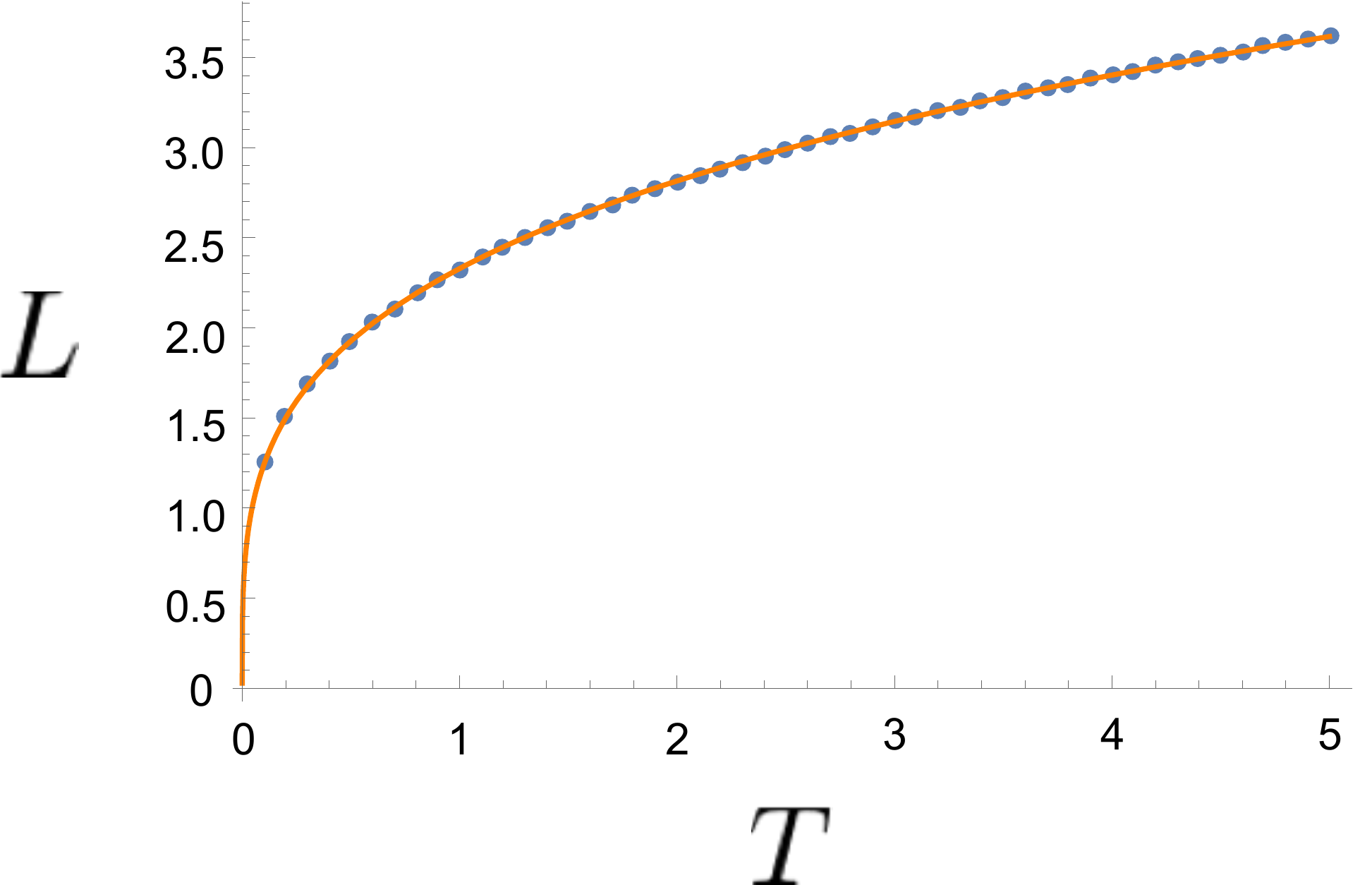}
		\caption{Points: Numerically calculated exponent $L$ for various values of the temperature $T$. Line: a fitting of the points by $\gamma T^c$ with constant $\gamma$ and $c$.
		}
		\label{fig:monomial_fit}
	\end{minipage}
\end{figure}

In what follows, we numerically calculate \eqref{OTOC in integral form} and investigate its temperature dependence, to show that the assumption made in Sec.~\ref{quantum Lyapunov} is a reasonable one. In the first place, we need to find the density of states $\rho(E)$ for calculating \eqref{OTOC in integral form}.
For that purpose, we define $W(E)$ which is the number of states below a given value $E$ of the energy. 
We numerically solve the Schr\"odinger equation for the CHO Hamiltonian system up to $E\simeq40$. Fig.~\ref{fig:the number of states} Left shows the numerically calculated number of states $W(E)$. To extract analytically the energy dependence of $W(E)$,  we fit it by a function of the form $W(E)=a+bE^c$. The fitting gives 
\begin{align}
\label{eq:the number of states}
	W(E) = -1.08 + 0.64E^{1.72}\, .
\end{align}
This function actually fit $W(E)$ nicely, as shown in Fig.~\ref{fig:the number of states} Right in which the continuous function \eqref{eq:the number of states} is overlaid  on the point plots of Fig.~\ref{fig:the number of states} Left. Differentiating \eqref{eq:the number of states} with respect to energy $E$, we find the density of states $\rho(E)$ as
\begin{align}
\label{density of states}
	\rho(E) = 1.10E^{0.72}\, .
\end{align}

Using this, the thermal OTOC \eqref{OTOC in integral form} is\footnote{Note that \eqref{density of states} is valid only for $E\lesssim40$ where the Schr\"odinger equation is solved. With \eqref{density of states}, we can deal with the thermal OTOCs at a temperature value at which the states with $E\lesssim40$ dominate. In Sec.~\ref{sec:exponential growth of OTOC}, we numerically calculated the thermal OTOC for $T\leq5$. When $T=5$, the Boltzmann factor $e^{-\beta E}$ for the energy $E=40$ is $e^{-8}\simeq0.00034$, which is small enough. Thus, when $T \leq 5$, the contribution of the energy levels with $E\geq40$ is well suppressed. In the numerical calculation of \eqref{OTOC in integral form}, we consider $T \leq 5$ and ignore $E \geq E_{\rm{trunc}}=40$. }
\begin{align}
\label{OTOC in integral form 2}
	C_T(t) \propto \frac{1}{Z(T)}\int_0^{E_{\rm{trunc}}=40} dE~ E^{0.72}e^{2aE^{1/4}t}e^{-\beta E}\, .
\end{align}
Next, we make the $E$ integration of \eqref{OTOC in integral form 2}, for which we need the normalization value $a$ of the classical Lyapunov exponent. We adopt the value $a=0.20$ obtained by a numerical evaluation of the classical Lyapunov exponent at $E=40$ (which is high
enough so that the system is totally chaotic as seen in Fig.~\ref{fig:Poincare section})\footnote{The classical Lyapunov exponent numerically depends on the initial conditions in the phase space. We will evaluate the effects of this deviation from $a=0.20$ later.}.

\begin{figure}[t]
\centering
	\includegraphics[width=80mm]{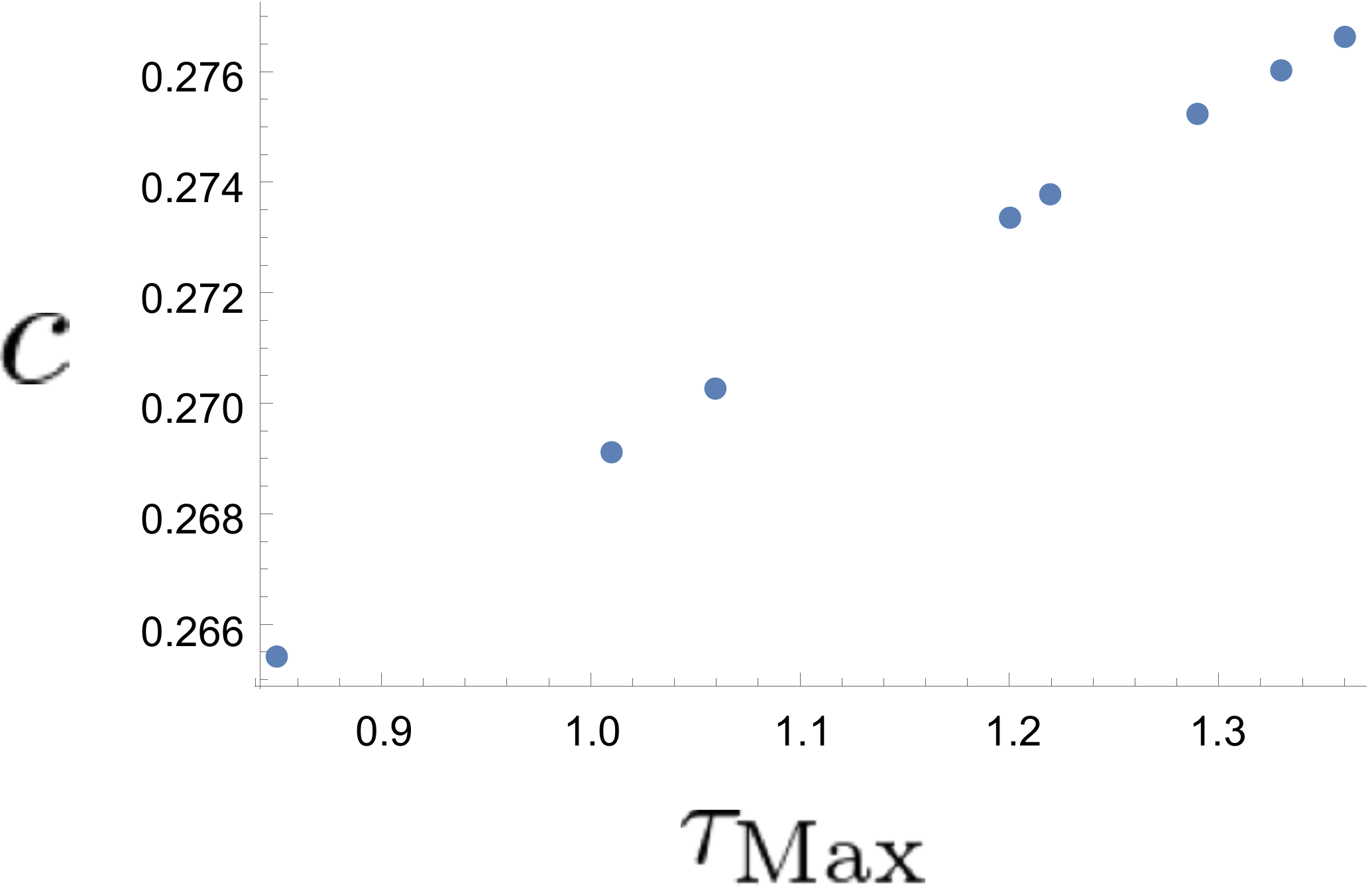}
	\caption{The exponent $c$ as a function of $\tau_{\rm{Max}}$. Each $\tau_{\rm{Max}}$ corresponds to a different initial condition for the numerical calculation of the classical Lyapunov exponent.
	}
	\label{fig:exponent}
\end{figure}

Considering the time domain $t \leq t_{\rm{Max}} = 6$ which is our expected Ehrenfest time observed in Fig.~\ref{fig:thermal OTOC}, and defining a rescaled time variable $\tau \equiv a t$
(and $\tau_{\rm{Max}} = at_{\rm{Max}} = 1.20$ accordingly), we perform the integral \eqref{OTOC in integral form 2} by discretizing\footnote{We discretize the variable $\tau$ by units of 0.01: $\tau = 0.01m~(m=0, 1, \cdots, m_{\rm{Max}}=120)$, and the temperature $T$ by units of 0.1: $T = 0.1n~(n=0, 1, \cdots, 50)$.} possible values of $\tau$ and $T$.

The numerical result $C_T(\tau)$ of \eqref{OTOC in integral form 2} at each fixed $T$ is fit by an exponential function $\alpha e^{L \tau}$, see Fig.~\ref{fig:exp_fit}.
We find\footnote{Note that this $L$ is different from what we call the quantum Lyapunov exponent by the time-rescaling factor $a$. However, since our target is just the power of $T$ in the quantum Lyapunov exponent, we don't need to worry about the overall rescaling factor.} the exponent $L$ as a function of $T$, see Fig.~\ref{fig:monomial_fit}. Then finally we numerically fit this set of $L$'s by a power function of $T$, $\gamma T^c$ with constants $c$ and $\gamma$, as shown in Fig.~\ref{fig:monomial_fit}. The result is: 
\begin{align}
\label{expectation}
	\lambda_{\rm{q}}(T) \propto T^c, \quad c\simeq 0.27\, .
\end{align}
This is consistent with the temperature dependence \eqref{temperature dependence} found in Sec.~\ref{quantum Lyapunov}.
And this confirms that the assumption we made in Sec.~\ref{quantum Lyapunov} is reasonable.

Our numerical result of the exponent $c$ in \eqref{expectation} depends on our truncation scheme and also on $a$, the normalization of the classical Lyapunov exponent. To evaluate possible statistical error due to the latter, we perform the numerical Lyapunov analysis for some different initial conditions at $E=40$, and find 
a distribution: $a=0.141, 0.169, 0.177, 0.200, 0.203,$ $0.215, 0.233, 0.226$.
This difference in $a$ results in the difference in $\tau_{\rm Max}$ as we set $t_{\rm{Max}}=6$. 
Then we numerically find that the exponent $c$ takes slightly different values depending on $\tau_{\rm{Max}}$, see Fig.~\ref{fig:exponent}. 
The resultant $c$ is distributed in a rather small range $0.26<c<0.28$. So we conclude that the result \eqref{expectation} is trustable against the errors in evaluation
of the classical Lyapunov exponent.


\section{Comparison to energy level statistics}
\label{sec:comparison}

\begin{figure}[t]
	\centering
		\includegraphics[width=70mm]{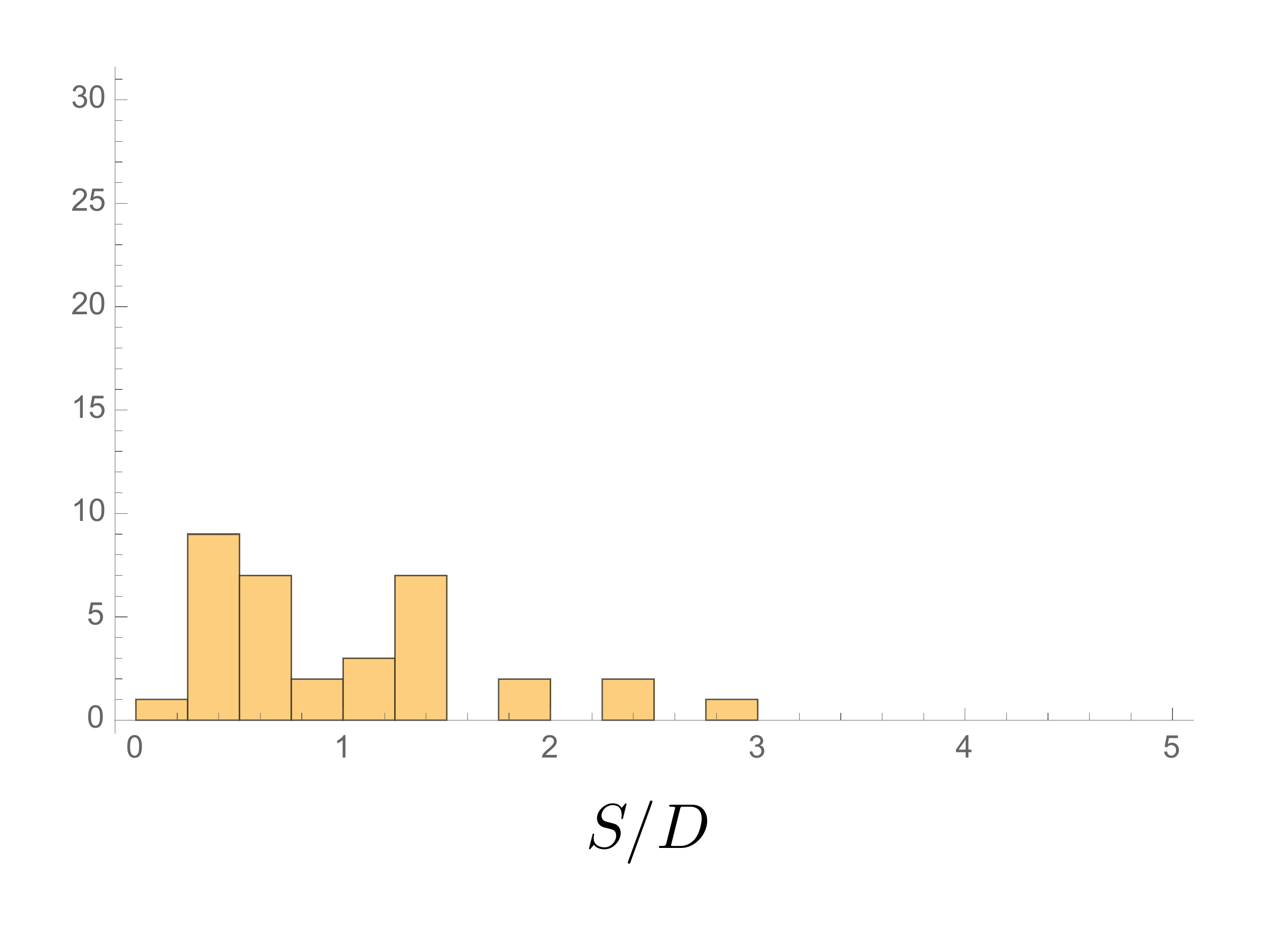}
		\hspace{3mm}
		\includegraphics[width=70mm]{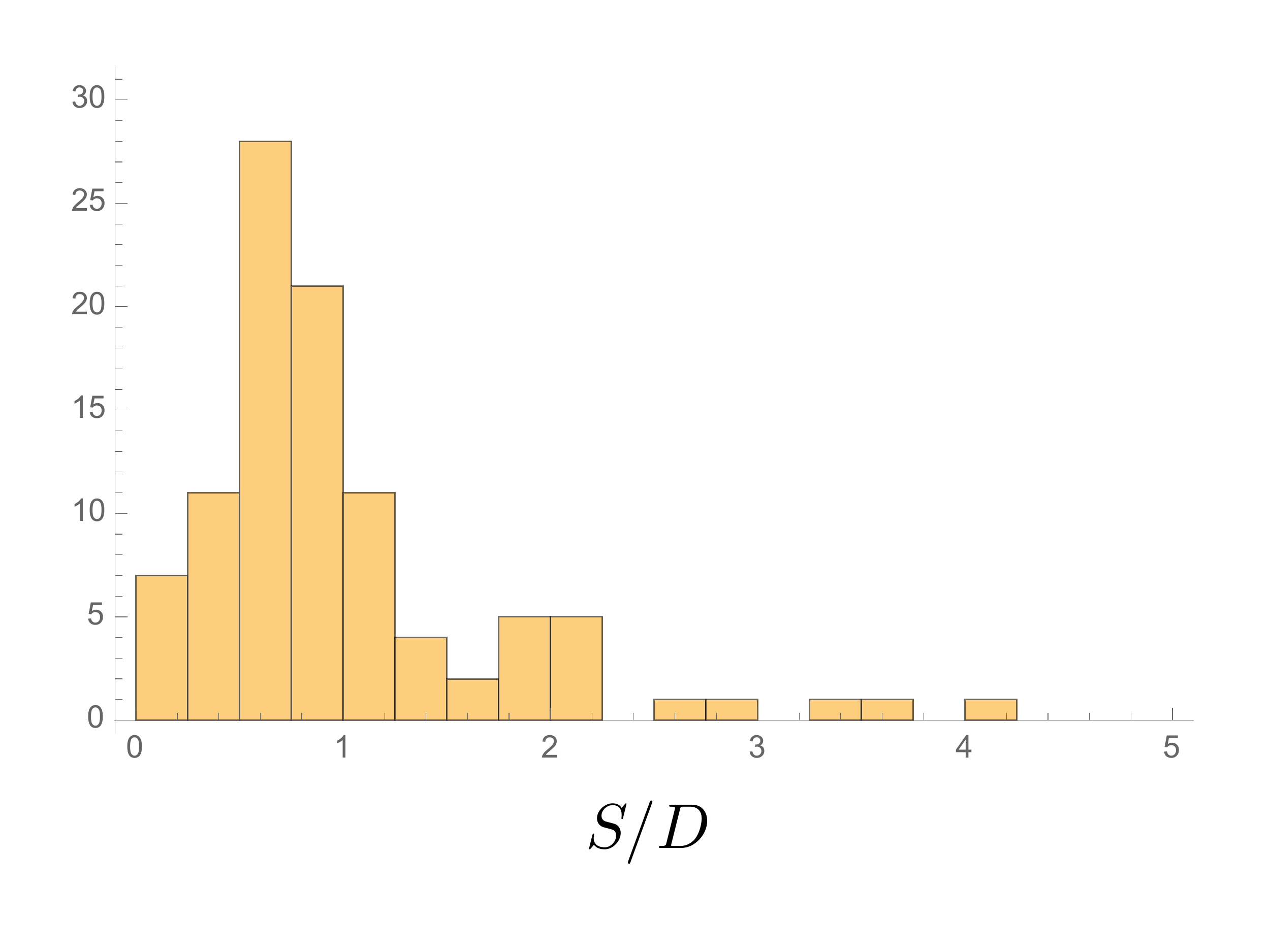}
		\caption{Histograms for the nearest-neighbor spacings of energy eigenvalues from 
		$E^{\rm (sym)}_{1}$ to $E^{\rm (sym)}_{35}$ (Left) 
		and $E^{\rm (sym)}_{1}$ to $E^{\rm (sym)}_{100}$
		 (Right)~. The horizontal axis is the nearest-neighbor spacing $S$ normalized by its
		 average value $D$ over the chosen energy interval.
		}
		\label{fig:level_statistics_from_1_to_35}
\end{figure}

\label{sec:level statistics}
So far, we have investigated the OTOC of the CHO model \eqref{CHO}. In Sec.~\ref{sec:exponential growth of OTOC} we found the exponential growth of the thermal OTOC, which indicates the quantum chaos, for the temperature range $3\leq T \leq 5$. On the other hand, as we mentioned in Sec.~\ref{subsec:reduction}, the distribution of the nearest-neighbor spacings of the energy eigenvalues is often used to discriminate chaoticity of the system. Regular systems show a Poisson distribution, while chaotic systems show a Wigner one. In the quantum analysis \cite{Haller Koppel Cederbaum (1984)}, the distribution of the energy eigenvalues of the CHO model was shown to be Wigner-like (Poisson-like) at high (low) energy. 

We are interested in whether our thermal OTOC is a better indicator of the quantum chaos compared to the energy level statistics. 
The analysis made in  \cite{Haller Koppel Cederbaum (1984)} used the CHO model with parameters different from ours. So, for the comparison let us 
find a relation between the parameters of ours and those of \cite{Haller Koppel Cederbaum (1984)}. 
Our Schr\"odinger equation is given in \eqref{Schr} with $\omega=1$ and $g_0=1/10$, while that of \cite{Haller Koppel Cederbaum (1984)} is 
\begin{align}
	-\frac12 \left(\frac{\partial^2}{\partial x^2}+\frac{\partial^2}{\partial y^2}\right) \psi_n(x, y) 
	+ \left[ \frac{1}{2} (x^2+y^2) + 4k x^2 y^2 \right] \psi_n(x, y) = \tilde{E}_n \psi_n(x, y) \, .
	\label{Schr2}
\end{align}
The numerical calculations in \cite{Haller Koppel Cederbaum (1984)} were made with the choice $k=1/200$.
When the nonlinear term is dominant and the system is totally chaotic, we may ignore the harmonic potential terms\footnote{If we did not ignore the
harmonic potential term, the exact comparison between our \eqref{Schr} and \eqref{Schr2} could not be made. So, here, we adopt that approximation.
We expect that even with this approximation the order of estimate is still valid.}. 
In that approximation, for \eqref{Schr2} to be identical to \eqref{Schr}, we need to make the rescaling $x \to a x$ and $b\tilde{E} = E$, with 
$a= ( 5/2)^{1/6}$ and $b=2^{2/3}5^{1/3}$. So we find the relation between the energy $E$ of our system and the energy $\tilde{E}$ of 
\cite{Haller Koppel Cederbaum (1984)},
\begin{align}
E = 2^{2/3} 5^{1/3} \tilde{E} \, .
\end{align}
In the analysis of \cite{Haller Koppel Cederbaum (1984)}, the energy level spacings show the Wigner distribution for the energy $\tilde{E} \geq 85$, while show
the Poisson one for $\tilde{E} \leq 85$.
Translated to our energy, this threshold energy scale is $E\lesssim 230$.
Now, our energy levels used for the evaluation of the thermal OTOC is $E \lesssim 32$. This means that the detection of the chaos by means of the OTOCs
is much more effective compared to the statistics of the energy level spacings.

To make sure that the argument above is correct, here we explicitly analyze the energy level spacings of our system \eqref{CHO}.
The statistical analysis needs to take care of the symmetry of the system, which is the point group $C_{4v}$ in our case. 
We consider wave functions which are completely symmetric under the
$C_{4v}$ group.  
First, to compare our OTOC analyses, we consider the energy eigenstates which we used
for the OTOC analyses: $E \lesssim 32$ (see Appendix \ref{sec:truncation error} for the truncation scheme).
Fig.~\ref{fig:level_statistics_from_1_to_35} Left is the obtained histogram for the nearest-neighbor spacings of energy eigenstates with the eigenvalue $\left\{ E^{\rm (sym)}_{n} \right\}$ 
($\lesssim 32$), 
whose wave functions are symmetric under $C_{4v}$. Those are 35 states among those shown in Fig.~\ref{energy_level}. The horizontal axis represents the nearest-neighbor spacing $S$ renormalized by its average value 
$D = (E^{\rm (sym)}_{35}-E^{\rm (sym)}_{1})/(35-1)$.
It is obvious that 
we lack the number of states sufficient to judge whether the histogram is subject to
the Wigner distribution or not.


To see the distribution, we need more states. If we use 100 states which are symmetric under $C_{4v}$ (whose highest energy goes up to $E \sim 60$), we obtain a histogram shown in 
Fig.~\ref{fig:level_statistics_from_1_to_35} Right. The horizontal axis is again normalized by 
$D = (E^{\rm (sym)}_{100}-E^{\rm (sym)}_{1})/(100-1)$.
This figure shows that the distribution is getting closer to the Wigner distribution. 

Therefore, we conclude that by the level statistics it is almost impossible 
to see the Wigner distribution only with the states which we used for the evaluation of the OTOCs.

Quantitatively, we can evaluate whether a given distribution is Wigner-like or not by the $\tilde{r}$-parameter \cite{Oganesyan Huse (2007), Atas Bogomolny Giraud Roux (2013)}, which is defined as
\begin{align}
\label{r-parameter}
	\tilde{r}_n = \frac{\min(s_n,s_{n-1})}{\max(s_n,s_{n-1})}, \quad s_n = E^{\rm (sym)}_n - E^{\rm (sym)}_{n-1}\, .
\end{align}
By definition, $\tilde{r}_n$ is a real number between 0 and 1. The distribution is characterized by the average of $\tilde{r}$-parameter, which we denote as $\langle \tilde{r} \rangle$. For example, $\langle \tilde{r} \rangle$ is around $0.386$ for a Poisson distribution. For a Wigner distribution, $\langle \tilde{r} \rangle$ takes a value in $0.531 \sim 0.674$. 
For the 35 energy eigenvalues in Fig.~\ref{energy_level} Left, we find
\begin{align}
	\langle \tilde{r} \rangle \simeq 0.478\, ,
\end{align}
which is different from that of a Wigner distribution. 
For the 100 energy eigenvalues in Fig.~\ref{energy_level} Right, we find
\begin{align}
	\langle \tilde{r} \rangle \simeq 0.526\, ,
\end{align}
which is closer to the region for the Wigner distribution.

Thus, the distribution of nearest-neighbor spacings of the energy eigenvalues which we used for the calculation of the OTOC is not 
sufficient to judge
a Wigner distribution. In other words, we cannot discriminate quantum chaoticity of the CHO model by the level statistics in the energy scale. In spite of this, we have observed the exponential growth of the thermal OTOC in Sec.~\ref{sec:exponential growth of OTOC}. We can conclude that the OTOC is more sensitive to quantum chaoticity than the level statistics, and is a better indicator of quantum chaos.

\begin{figure}[t]
\centering
	\includegraphics[width=130mm]{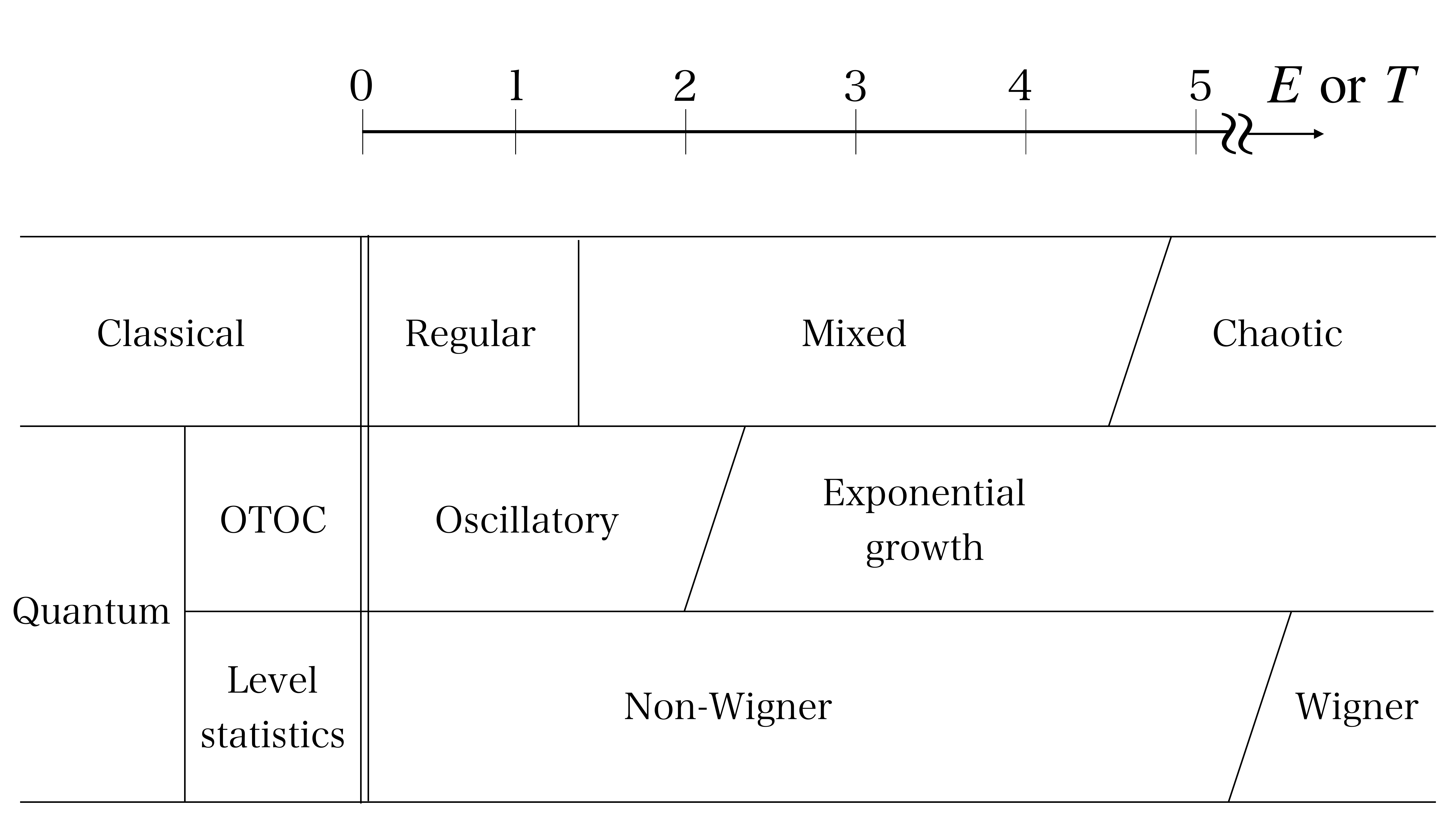}
	\caption{Summary table of the energy (temperature) dependence of the classical/quantum chaos of the CHO system \eqref{CHO}.
	The thermal OTOC can detect the quantum chaos at lower energy (temperature) compared with the level statistics. The quantum regular/chaos phase structure through the OTOC is found to be consistent with the classical phase structure, which indicates that the thermal OTOC properly reflects the property of the classical system.
	}
	\label{fig:summary}
\end{figure}

\section{Conclusion and discussion}
\label{sec:conclusion}

In this paper, 
we numerically calculated the microcanonical OTOC and the thermal OTOC of the nonlinearly coupled harmonic oscillator \eqref{CHO}, for various temperature and energy range. 
The nonlinear coupling is a reduction of SU(2) Yang-Mills theory which is expected to have a gravity dual in the 
large $N$ and strong coupling limit. 
We confirmed the exponential growth of the thermal OTOC at higher temperature, and could read the quantum Lyapunov exponent. 
The Lyapunov exponent measured has a temperature dependence $\lambda_{\rm q} \propto T^c$ with $c =0.26 \sim 0.31$. 

We have two reasons for the identification of the quantum Lyapunov exponent. 
First, the order of magnitude of the calculated exponent (Fig.~\ref{fig:quantum Lyapunov}) coincided with that of the classical Lyapunov exponent, $\lambda_{\rm{cl}}$ (Fig.~\ref{fig:classical Lyapunov}), as shown in Sec.~\ref{sec:exponential growth of OTOC}.
Second, the temperature dependence of the calculated exponent \eqref{temperature dependence} reproduces the dependence \eqref{expectation} expected from the energy dependence of the classical Lyapunov exponent, $\lambda_{\rm{cl}}$, \eqref{energy dependence}, as shown in Sec.~\ref{energy and temperature}.

In Sec.~\ref{sec:level statistics} we showed that the level statistics of the energy eigenstates that are effective for the calculation of the OTOC does not show a Wigner distribution. This means that the level statistics fails in discriminating quantum chaos of the coupled harmonic oscillator model at the energy scale of our concern. 
On the other hand, we found the exponential growth of the thermal OTOC. Therefore the thermal OTOC is a better indicator of quantum chaos than the level statistics, for the nonlinearly coupled harmonic oscillator system.

Fig.~\ref{fig:summary} summarizes our quantum results and their relationship with the classical regular/chaos phase structure. The horizontal axis is the temperature (or energy) of the quantum and classical systems. 
In the classical coupled harmonic oscillator system described in Sec.~\ref{subsec:classical properties}, at low energy, we see the orbits in the Poincar\'e section, so the system is in a regular phase, while at high energy, orbits decay and the Poincar\'e section is filled with scattered points, meaning that the system is in a chaos phase.
In the quantum system, the thermal OTOC is more sensitive to the quantum chaos than the level statistics, and is able to discriminate the quantum chaos at lower energy (temperature).
Moreover, the ``phase transition'' temperature between the OTOCs oscillatory phase and exponential growth phase is observed to close to the 
energy at which the classical regular/chaos phase transition occurs. 
Thus the thermal OTOC properly reflects the chaotic property of the classical system.

It is encouraging that even the very simple coupled harmonic oscillators exhibit the exponential growth of the thermal OTOC.
Inclusion of more oscillators will make the Ehrenfest time later, which will help reading the quantum Lyapunov exponent.
For the system to get closer to the black hole system, according to the generic AdS/CFT dictionary, 
we need to take the large $N$ limit and the strong coupling limit. For the latter we need to go to lower temperature
because the coupling constant has a positive mass dimension, and thus need also to suppress the harmonic oscillator frequency $\omega$ 
to have the chaotic phase at lower energy scales.
Achieving these limits will enables us to find the quantum Lyapunov exponent saturating
the bound $2\tilde{\lambda}_{\rm q}=2\pi T$, the black hole behavior. 
The emergence of this truly quantum regime would be quite intriguing for understanding quantum gravity.

Also in the introduction we emphasized the relation to the holographic principle, that is why we considered the thermal OTOCs.
Another motivation for using the thermal OTOC is that it captures a global feature of the phase space.
Classical instability could be easily captured by localized wave functions with which the expectation value of the OTOC is taken.
However, if one localizes the wave function near the top of the potential hill, then even when the whole system is not chaotic,
the OTOC can grow exponentially (see for example \cite{Pilatowsky-Cameo:2019qxt}). 
The local growth does not immediately mean chaos, and in fact,
this kind of exponential growth can occur even in an inverted harmonic oscillator \cite{Ali:2019zcj}
or in a double-well potential \cite{Bzowski:2018aiq}
in one dimension.
The thermal OTOCs, on the other hand, use energy eigenstates which are spread over the whole phase space,
thus would not see these local instabilities. 
Since ergodicity is expected for classically chaotic systems,
even widely spread wave functions for the thermal OTOCs can detect the chaos due to the global structure of the phase space.

One of the importance of our system \eqref{CHO} is its genericity. Harmonic oscillators are everywhere, as a basis of quantum field theories
and quantum information theories. They are also easier to be implemented as experimental apparatus, and at finite temperature the
thermal OTOC is appropriate to serve as a better diagnosis for finding quantum chaos. We expect that some harmonic oscillator
couplings other than our \eqref{CHO} will produce quantum chaos similarly, with different scaling properties. Studying the
thermal OTOCs and scrambling timescales
in systems given by some limits of coupled harmonic oscillators will open the way to explicitly examine 
quantum gravity nature of matters.


\acknowledgments

We would like to thank Toshihiro Ota for valuable discussions 
at the early stage of this work.
K.~H.~would like to thank Stefan Heusler, 
Keiju Murata, Lea Ferreira dos Santos, and Ryosuke Yoshii 
for valuable discussions.
This work is supported in part by JSPS KAKENHI Grant No.~JP17H06462.


\appendix
\section{Evaluation of the level truncation error}
\label{sec:truncation error}

In this appendix, we evaluate possible systematic errors coming from the level truncation of the energy eigenstates, 
and show that our main statements in this paper do not depend on the truncation error.   

\begin{figure}[t]
\centering
	\includegraphics[width=100mm]{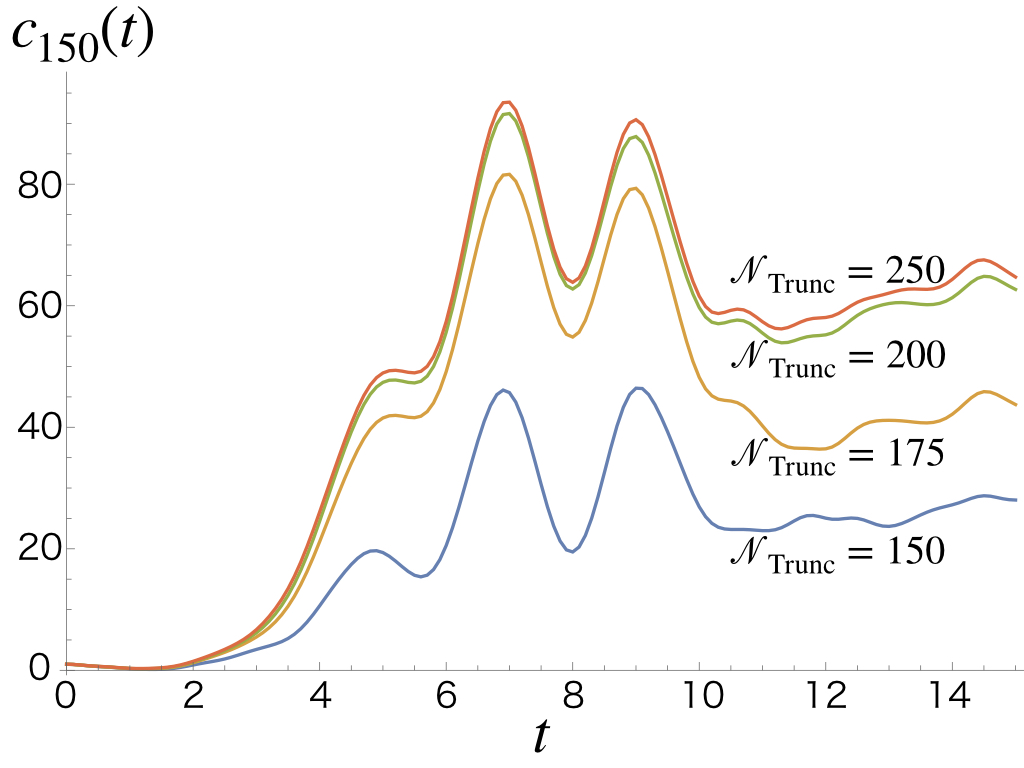}
	\caption{Microcanonical OTOC $c_n(t)$ with $n=150$ given in \eqref{microcanonical}, with the sum over the index $m$ truncated at $\mathcal{N}_{\rm{trunc}}$.}
	\label{fig:truncation}
\end{figure}

As described in Sec.~\ref{subsec:numerics}, the microcanonical and the thermal OTOCs are calculated by infinite series. For the numerical calculation, the infinite series must be truncated.  
First, we look at the microcanonical OTOC defined in \eqref{microcanonical}. We truncate the sum over the index $m$ of \eqref{microcanonical} at $\mathcal{N}_{\rm{trunc}}$, and investigate the $\mathcal{N}_{\rm{trunc}}$ dependence of the microcanonical OTOC. In Fig.~\ref{fig:truncation}, we plot $c_{150}(t)$ for $\mathcal{N}_{\rm{trunc}}=150,175,200,250$. 
We find a good convergence for $\mathcal{N}_{\rm{trunc}}=250$. 

Let us consider the evaluation of the thermal OTOC. 
At the highest value of the temperature $T = 5$ in our analysis, the contribution of the microcanonical OTOC $c_n$ with $n>150$ to the thermal OTOC in negligible. This is because the Boltzmann factor $\exp \left( -E_n/T \right)$ in \eqref{thermal and microcanonical} is exponentially suppressed.
For example, at $T=5$, $c_{150}(t)$ contributes only with the weight $\exp \left( -E_{150}/5 \right) \approx 0.0085$ to $C_T(t)$. Thus the microcanonical OTOC with $n>150$ do not contribute the thermal OTOC $C_T(t)$ so much. 

Based on the considerations above, for computational reason, we choose $\mathcal{N}_{\rm{trunc}}=253$, and take 150 for the number of the microcanonical OTOCs $c_n(t)$ for calculating the thermal OTOC. 
This $\mathcal{N}_{\rm{trunc}}=253$ corresponds to the state with its energy $E\simeq 32$.


\end{document}